\documentclass[aps,preprint,showpacs,showkeys]{revtex4}
\usepackage{amsmath}
\usepackage{graphicx}
\usepackage{bm}

\begin{document}

\title{Plasmon oscillations in ellipsoid nanoparticles: beyond dipole
approximation}
\author{D.V. Guzatov$^{1}$}
\email{guzatov@gmail.com}
\author{V.V. Klimov$^{2}$}
\email{vklim@sci.lebedev.ru}
\author{M.Yu. Pikhota$^{2}$}
\affiliation{$^{1}$Research Center of Resource Saving Problems, NAS Belarus, Grodno
230023, Tyzenhauz sq. 7 \\
$^{2}$P.N. Lebedev Physical Institute, Russian Academy of Sciences, Moscow
119991, Leninsky prospect 53, ``Advanced Energy Technologies'', LTD}

\begin{abstract}
The plasmon oscillations of a metallic triaxial ellipsoid
nanoparticle have been studied within the framework of the
quasistatic approximation. A general method has been proposed for
finding the analytical expressions describing the potential and
frequencies of the plasmon oscillations of an arbitrary
multipolarity order. The analytical expressions have been derived
for an electric potential and plasmon oscillation frequencies of the
first 24 modes. Other higher orders plasmon modes are investigated
numerically.
\end{abstract}

\date{\today}
\pacs{41.20.Cv, 73.20.Mf, 78.67.-n}
\keywords{localized plasmons, triaxial ellipsoid, nanoantenna}
\maketitle

\section{Introduction}

Quite a number of works were recently devoted to the studies of optical
properties of nanodimensional bodies. Special attention has been paid to the
metallic nanoparticles, with the help of which one could amplify the near
electric field at the frequencies of the localized plasmon resonances \cite%
{ref1}. Possible applications of that effect have been considered. The most
developed of them is the use of large local fields near a rough surface for
increasing the surface-enhanced Raman scattering (SERS) \cite{ref2}.
Modification of properties of the radiating atoms by nanobodies of different
shape and composition has been the basis for research and development of
nanobiosensors \cite{ref3}-\cite{ref6}, nanolasers \cite{ref7}, microscopes
that are capable to detect single molecules \cite{ref8}, the devices for DNA
decoding \cite{ref9}, chemical sensors \cite{ref10}-\cite{ref12}, and many
others.

Very recently, the plasmon nanostructures were successfully used in optical
devices, in which the diffraction limit had been overcome thus providing an
effective interaction between near and far fields, e.g. in hyper- and
nanolenses \cite{ref13}-\cite{ref15}, nanoantennas \cite{ref16}-\cite{ref20}%
. Studying optical properties of the plasmon nanoparticles made up a new
line of investigation such as ``plasmonics'' or ``nanoplasmonics'' \cite%
{ref21,ref22}.

The above-mentioned applications are based on properties of single
nanoparticles or clusters of such nanoparticles. Today, the optical
properties of metallic nanospheres \cite{ref23}-\cite{ref30},
nanospheroids \cite{ref31}-\cite{ref34}, and some other nanobodies
\cite{ref21,ref22,ref35}, have been sufficiently well studied with
the help of known analytical approaches. Many of the other shapes of
nanoparticles have been studied only numerically, which is often
insufficient for understanding of underlying physics and for
development of new nanodevices.

It is astonishing but an important class of nontrivial nanoparticle
shapes , which can be described analytically, turns out to be almost
uninvestigated. We keep in mind nanoparticles of triaxial ellipsoid
shape. Such nanoparticles have very different shapes and are often
studied experimentally \cite{ref36}-\cite{ref39}. Despite a great
number of theoretical works \cite{ref40}-\cite{ref52} devoted to
that problem, there had not yet been achieved any appreciable
progress in studies of optical properties of triaxial ellipsoids.
This is due, first of all, to an extreme mathematical complexity of
the problem. In practice, to describe optical properties of triaxial
ellipsoid nanoparticles one makes use of a simple solution for a
nanoellipsoid in a uniform electric field \cite{ref53} that
describes only the simplest (``dipole'') plasmon oscillations. Such
a description is sufficient if scattering of plane electromagnetic
waves by nanoparticle is of interest. But for many nanooptical
problems there are no plane waves and such a solution cannot be
used. In particular, it does not work when considering an
ellipsoidal nanoparticle interaction with highly nonuniform fields,
which is fundamentally important for a nanoworld. For example, the
radiation fields of atoms, molecules, or quantum dots placed near
the nanobodies are essentially not uniform, and they effectively
excite not only ``dipole'', but more complex plasmon oscillations
also. For example, exactly these modes explain very important
phenomenon of radiation quenching.

The aim of this work is to study optical properties of a triaxial metallic
nanoellipsoid within the quasistatic approximation, in respect to the
excitation of high-order plasmon oscillations that are somehow analogous to
the quadrupole, octopole, and hexadecapole spherical harmonics. Here we
obtain explicit analytical expressions, which is especially important for
fast calculations and estimations.

The content of the remaining part of the article is as follows. In section
II, general approach to finding the eigen-functions of plasmon oscillation
electric field potential and the permittivity (frequencies) corresponding to
those oscillations is described. In section III, the above method is applied
to determine explicit values of the permittivity and the plasmon mode
potentials. In section IV, the graphic illustrations of the found solutions
are given and the results obtained are discussed. The geometry of the
problem in question is illustrated in Fig.1.

\section{General method of solution}

The localized plasmons naturally arise as the nontrivial solutions of a
homogeneous quasistatic problem, that is, the Laplace equations with the
corresponding boundary conditions \cite{ref54,ref55}

\begin{eqnarray}
\Delta \phi _{n} &=&0,  \notag \\
\Delta g_{n} &=&0,  \notag \\
\left. \varepsilon _{n}\frac{\partial \phi _{n}}{\partial \mathbf{n}}%
\right\vert _{S} &=&\left. \bar{\varepsilon}\frac{\partial g_{n}}{\partial
\mathbf{n}}\right\vert _{S},  \label{eq1}
\end{eqnarray}

\noindent where $\bar{\varepsilon}$ is the permittivity of space around a
nanoparticle that we take equal to 1; $\phi _{n}$, $g_{n}$, the plasmon
potentials inside and outside a nanoparticle, respectively; $\left. \frac{%
\partial \phi _{n}}{\partial \mathbf{n}}\right\vert _{S}$ stands for a
normal derivative of the plasmon potential $\phi _{n}$ at a particle
surface; $\varepsilon _{n}$, the resonant permittivity corresponding to the
plasmon oscillations of the order $n$, and $\Delta $, the Laplace operator.
The latter equation in (\ref{eq1}) stands for a fulfillment of continuity of
the induction normal components . One may show \cite{ref54,ref55} that a
homogeneous system (\ref{eq1}) has nontrivial solutions (localized plasmons)
for some (negative) parameters $\varepsilon _{n}$ only. \ It is naturally to
call $\varepsilon _{n}$ as resonant permittivities. If the permittivity of a
real particle is close to one of the values $\varepsilon _{n}$ then it will
have a plasmon resonance at frequency $\omega _{n}$, at which

\begin{equation}
\varepsilon _{n}\approx \varepsilon \left( \omega _{n}\right) .  \label{eq2}
\end{equation}

\noindent That frequency $\omega _{n}$ is called the plasmon frequency and
can be found if the dispersion law $\varepsilon \left( \omega \right) $ is
known. In what follows we will use the Drude's dispersion law, ${\varepsilon
\left( \omega \right) }=1-{\left( \omega _{pl}{/}\omega \right) }^{2}$,
where $\omega _{pl}$ is the bulk plasma frequency of nanoparticle material
(metal).

The eigenfunctions of the plasmons $\mathbf{e}_{n}\left( \mathbf{r}\right)
=-\nabla \phi _{n}\left( \mathbf{r}\right) $ (where $\nabla $ is the
gradient operator) are orthogonal in sense that ($n\neq m$)

\begin{equation}
\int\limits_{V}d\mathbf{r}\left( \mathbf{e}_{n}\cdot \mathbf{e}_{m}\right) =0
\label{eq3}
\end{equation}

\noindent These eigenfunctions make the complete system, and the solution
for a nanoparticle problem in external field $\mathbf{E}^{0}$ may be
expanded over \ wavefunctions \ $\mathbf{e}_{n}\left( \mathbf{r}\right) $ of
the \ localized plasmons \cite{ref55}

\begin{equation}
\mathbf{E}\left( \mathbf{r}\right) =\mathbf{E}^{0}\left( \mathbf{r}\right)
+\sum\limits_{n}\mathbf{e}_{n}\left( \mathbf{r}\right) \left( \frac{%
\varepsilon \left( \omega \right) -1}{\varepsilon _{n}-\varepsilon \left(
\omega \right) }\right) \frac{\int\limits_{V}d\mathbf{r}\left( \mathbf{e}%
_{n}\cdot \mathbf{E}^{0}\right) }{\int\limits_{V}d\mathbf{r}\left( \mathbf{e}%
_{n}\cdot \mathbf{e}_{n}\right) },  \label{eq4}
\end{equation}

\noindent where $\varepsilon \left( \omega \right) $ is the dielectric
constant of nanoparticle and the integration is performed over a
nanoparticle volume. Thus, this problem is reduced to finding the potentials
of plasmon modes and corresponding eigenvalues (resonant permittivities).

To determine the properties of localized plasmons in a triaxial
nanoellipsoid it is convenient to use an integral formulation, instead of
system (\ref{eq1}) (see, for instance, \cite{ref56})

\begin{equation}
\phi _{n}\left( \mathbf{r}\right) =\frac{\varepsilon _{n}-1}{4\pi }%
\int\limits_{V}d\mathbf{r}^{\prime }\left( \left( \nabla ^{\prime }\phi
_{n}\left( \mathbf{r}^{\prime }\right) \right) \cdot \nabla \right) \frac{{1}%
}{\left\vert \mathbf{r}^{\prime }-\mathbf{r}\right\vert },  \label{eq5}
\end{equation}

\noindent where the integration is performed over the nanoparticle volume; $%
\mathbf{r}^{\prime }$, $\mathbf{r}$ are the coordinates of some two points
inside the nanoparticle, and ${\nabla }^{\prime }$, $\nabla $, the operator
gradient over the coordinates $\mathbf{r}^{\prime }$, $\mathbf{r}$,
respectively. Because the $\mathbf{r}$ point is inside ellipsoid, expression
(\ref{eq5}) is the integral equation. The solution of (\ref{eq5}) gives us
both the eigenvalues $\varepsilon _{n}$ and the values of the plasmon
potentials at any spatial point inside ellipsoid.

Due to eq.(\ref{eq1}),the solution of integral equation (\ref{eq5}) is a
bounded harmonic function of ellipsoidal symmetry ( ellipsoidal harmonics).
The latter may be represented in the form of the Niven's functions \cite%
{ref57}-\cite{ref59} which are the polynomials of the Cartesian coordinates $%
\left\{ x_{1},x_{2},x_{3}\right\} $:

\begin{eqnarray}
\Phi \left( x_{1},x_{2},x_{3}\right) &=&\left\{
\begin{array}{cccc}
& x_{1} & x_{2}x_{3} &  \\
1 & x_{2} & x_{3}x_{1} & x_{1}x_{2}x_{3} \\
& x_{3} & x_{1}x_{2} &
\end{array}%
\right\}  \notag \\
&&\times \Theta _{1}\left( x_{1},x_{2},x_{3}\right) \Theta _{2}\left(
x_{1},x_{2},x_{3}\right) ...\Theta _{p}\left( x_{1},x_{2},x_{3}\right) ,
\label{eq6}
\end{eqnarray}

\noindent where

\begin{equation}
\Theta _{p}\left( x_{1},x_{2},x_{3}\right) =\frac{x_{1}^{2}}{%
a_{1}^{2}+\theta _{p}}+\frac{x_{2}^{2}}{a_{2}^{2}+\theta _{p}}+\frac{%
x_{3}^{2}}{a_{3}^{2}+\theta _{p}}-1.  \label{eq7}
\end{equation}

\noindent In Eq.(\ref{eq7}), $\theta _{p}$ is a constant. By using the
ellipsoid coordinates \cite{ref57} it is easy to prove that the Niven's
functions (\ref{eq6}) are the eigenpotentials of plasmons inside an
ellipsoid.

Thus, knowing the eigenfunctions (\ref{eq6}) of eqs. (\ref{eq1}) and (\ref%
{eq5}) one can represent the corresponding eigenvalues ( resonant values of
permittivity) \ in the following form

\begin{equation}
\varepsilon _{n}=1+\left\{ \frac{1}{4\pi \phi _{n}\left( \mathbf{r}\right) }%
\int\limits_{V}d\mathbf{r}^{\prime }\left( \left( \nabla ^{\prime }\phi
_{n}\left( \mathbf{r}^{\prime }\right) \right) \cdot \nabla \right) \frac{1}{%
\left\vert \mathbf{r}^{\prime }-\mathbf{r}\right\vert }\right\} ^{-1},
\label{eq8}
\end{equation}

\noindent and the problem is thereby reduced to calculating the
integral of the Niven's functions (\ref{eq6}) over a nanoellipsoid
volume. The integration can be made by using the Dyson's theorem
\cite{ref60,ref61} (see Appendix A). According to it, an integral
over an ellipsoid volume of a
polynomial times Green function $1/\left\vert \mathbf{r}^{\prime }-\mathbf{r}%
\right\vert $, proves to be the polynomial as well, whose power is greater
by two

\begin{equation}
\int\limits_{V}d\mathbf{r}^{\prime }\frac{x_{1}^{\prime }{}^{p}x_{2}^{\prime
}{}^{q}x_{3}^{\prime }{}^{s}}{\left\vert \mathbf{r}^{\prime }-\mathbf{r}%
\right\vert }=P_{p+q+s+2}\left( x_{1},x_{2},x_{3}\right) ,  \label{eq9}
\end{equation}

\noindent where $P_{p+q+s+2}\left( x_{1},x_{2},x_{3}\right) $ is the
polynomial of the Cartesian coordinates ($\sum\limits_{\alpha =1}^{3}\left(
x_{\alpha }/a_{\alpha }\right) ^{2}<1$) of the power $p+q+s+2$.

To find a solution for the plasmonic oscillation potential outside an
ellipsoid one can use again (\ref{eq5})

\begin{equation}
g_{n}\left( r\right) =\frac{\varepsilon _{n}-1}{4\pi }\int\limits_{V}d%
\mathbf{r}^{\prime }\left( \left( \nabla ^{\prime }\phi _{n}\left( \mathbf{r}%
^{\prime }\right) \right) \cdot \nabla \right) \frac{1}{\left\vert \mathbf{r}%
^{\prime }-\mathbf{r}\right\vert }.  \label{eq10}
\end{equation}

\noindent\ but in eq.(\ref{eq10}), unlike (\ref{eq5}), $\mathbf{r}$ are
coordinates of the point outside the ellipsoid. Knowing the electric
potential of plasmon oscillations inside and outside a nanoellipsoid, as
well as the corresponding values of the permittivity $\varepsilon _{n}$, one
can find electric fields near and inside ellipsoidal nanoparticles in an
arbitrary external field (\ref{eq4}).

\section{Functions of electric potential and values of permittivity
corresponding to plasmon oscillations of a triaxial nanoellipsoid}

In this section, we will apply our general method \ to find the plasmonic
modes of the power $n=1,2,3,4$.

\begin{center}
\textbf{The case $n=1$ ``dipole'' modes}
\end{center}

Here, in accordance with (\ref{eq6}), the potential inside the ellipsoid is
linear over the coordinates

\begin{equation}
\phi _{1}=\sum\limits_{\alpha =1}^{3}A_{1}^{\left( \alpha \right) }x_{\alpha
}.  \label{eq11}
\end{equation}

\noindent By substituting (\ref{eq11}) into (\ref{eq5}), and making
the required differentiation and integration by the Dyson's theorem
(see Appendix A) we obtain the following system of equations:

\begin{equation}
A_{1}^{\left( \alpha \right) }=-\frac{1}{2}\left( \varepsilon _{1}-1\right)
a_{1}a_{2}a_{3}I_{\alpha }A_{1}^{\left( \alpha \right) },  \label{eq12}
\end{equation}

\noindent where $I_{\alpha }=\int\limits_{0}^{\infty }\frac{du}{\left(
u+a_{\alpha }^{2}\right) R\left( u\right) }$ and $R\left( u\right)
=\prod\limits_{\alpha =1}^{3}\sqrt{u+a_{\alpha }^{2}}$. As follows from (\ref%
{eq12}), there are three different eigen-functions ($\alpha =1,2,3$)
describing the electric potential of the plasmonic oscillations

\begin{equation}
\phi _{1}^{\left( \alpha \right) }=x_{\alpha },  \label{eq13}
\end{equation}

\noindent and three \ corresponding values of the permittivity

\begin{equation}
\varepsilon _{1}^{\left( \alpha \right) }=1-\left( \frac{a_{1}a_{2}a_{3}}{2}%
I_{\alpha }\right) ^{-1}.  \label{eq14}
\end{equation}

\noindent The found eigenfunctions (\ref{eq13}) and the eigenvalues (\ref%
{eq14}) are well known and widely used for description of the plasmon
resonances in ellipsoid nanoparticles in uniform field \cite{ref53}.

The electric potential of the plasmons with $n=1$ outside the
particle can be found from (\ref{eq10}) by using (\ref{eq13}) and
(\ref{eq14}) (see Appendix B).

\begin{center}
\textbf{The case $n=2$ ``quadrupole'' modes}
\end{center}

Here, the ellipsoid potential is described by the Niven's functions of two
types. Functions of the first type are proportional to $\Theta _{1}\left(
x_{1},x_{2},x_{3}\right) $, that is, can be represented in the form:

\begin{equation}
\phi _{2}^{\left( {1}\right) }=\sum\limits_{\beta =1}^{3}\frac{x_{\beta }^{2}%
}{\Lambda ^{\prime }-a_{\beta }^{2}}+1,\quad \phi _{2}^{\left( {2}\right)
}=\sum\limits_{\beta =1}^{3}\frac{x_{\beta }^{2}}{\Lambda ^{\prime \prime
}-a_{\beta }^{2}}+1,  \label{eq15}
\end{equation}

\noindent where ${\Lambda }^{\prime }$ and ${\Lambda }^{\prime \prime }$ are
the roots of a quadratic equation $\sum\limits_{\alpha =1}^{3}\frac{1}{%
\Lambda -a_{\alpha }^{2}}=0$, that is,

\begin{equation}
\left\{
\begin{array}{c}
\Lambda ^{\prime } \\
\Lambda ^{\prime \prime }%
\end{array}%
\right\} =\frac{1}{3}\left\{ \pm \left[ \sum\limits_{\alpha =1}^{3}\left(
a_{\alpha }^{4}-\frac{a_{1}^{2}a_{2}^{2}a_{3}^{2}}{a_{\alpha }^{2}}\right) %
\right] ^{1/2}+\sum\limits_{\alpha =1}^{3}a_{\alpha }^{2}\right\} .
\label{eq16}
\end{equation}

\noindent By substituting (\ref{eq15}) into (\ref{eq8}) and calculating the
emerging integrals we obtain explicit expressions for the permittivity of
those modes:

\begin{eqnarray}
\varepsilon _{2}^{\left( 1\right) } &=&1+\left( \frac{a_{1}a_{2}a_{3}}{2}%
\sum\limits_{\alpha =1}^{3}\frac{a_{\alpha }^{2}I_{\alpha }}{\Lambda
^{\prime }-a_{\alpha }^{2}}\right) ^{-1},  \notag \\
\varepsilon _{2}^{\left( 2\right) } &=&1+\left( \frac{a_{1}a_{2}a_{3}}{2}%
\sum\limits_{\alpha =1}^{3}\frac{a_{\alpha }^{2}I_{\alpha }}{\Lambda
^{\prime \prime }-a_{\alpha }^{2}}\right) ^{-1}.  \label{eq17}
\end{eqnarray}

Another type of the solutions for the plasmonic modes with $n=2$ includes
three Niven's functions

\begin{equation}
\phi _{2}^{\left( 3\right) }=x_{2}x_{3},\quad \phi _{2}^{\left(
4\right) }=x_{3}x_{1},\quad \phi _{2}^{\left( 5\right) }=x_{1}x_{2}.
\label{eq18}
\end{equation}

\noindent By substituting (\ref{eq18}) into (\ref{eq5}), and using the
Dyson's theorem one can find the corresponding values of the dielectric
permittivity:

\begin{eqnarray}
\varepsilon _{2}^{\left( 3\right) } &=&1-\left( \frac{a_{1}a_{2}a_{3}}{2}%
\left( a_{2}^{2}+a_{3}^{2}\right) I_{23}\right) ^{-1},  \notag \\
\varepsilon _{2}^{\left( 4\right) } &=&1-\left( \frac{a_{1}a_{2}a_{3}}{2}%
\left( a_{3}^{2}+a_{1}^{2}\right) I_{31}\right) ^{-1},  \notag \\
\varepsilon _{2}^{\left( 5\right) } &=&1-\left( \frac{a_{1}a_{2}a_{3}}{2}%
\left( a_{1}^{2}+a_{2}^{2}\right) I_{12}\right) ^{-1},  \label{eq19}
\end{eqnarray}

\noindent where $I_{\alpha \beta }=\int\limits_{0}^{\infty }\frac{du}{\left(
u+a_{\alpha }^{2}\right) \left( u+a_{\beta }^{2}\right) R\left( u\right) }$ (%
$\alpha ,\beta =1,2,3$).

The plasmonic oscillation potentials outside nanoparticle can again
be found by using (\ref{eq10}) (see Appendix B).

\begin{center}
\textbf{The case $n=3$ ``octupole'' modes}
\end{center}

In this case, the ellipsoid potential is also described by the Niven's
functions of two types. Functions of the first type have the form ($\alpha
=1,2,3$)

\begin{equation}
\phi _{3}^{\left( 2\alpha -1\right) }=x_{\alpha }\left( \sum\limits_{\gamma
=1}^{3}\frac{x_{\gamma }^{2}}{\Lambda _{\alpha }^{\prime }-a_{\gamma }^{2}}%
+1\right) ,\quad \phi _{3}^{\left( 2\alpha \right) }=x_{\alpha }\left(
\sum\limits_{\gamma =1}^{3}\frac{x_{\gamma }^{2}}{\Lambda _{\alpha }^{\prime
\prime }-a_{\gamma }^{2}}+1\right) .  \label{eq20}
\end{equation}

\noindent where $\Lambda _{\alpha }^{\prime }$ and $\Lambda _{\alpha
}^{\prime \prime }$ are the roots of the quadratic equation $%
\sum\limits_{\beta =1}^{3}\frac{1+2\delta _{\alpha \beta }}{\Lambda _{\alpha
}-a_{\beta }^{2}}=0$, that is,

\begin{eqnarray}
\left\{
\begin{array}{c}
\Lambda _{\alpha }^{\prime } \\
\Lambda _{\alpha }^{\prime \prime }%
\end{array}%
\right\} &=&\frac{1}{5}\left\{ \pm \left[ 4\sum\limits_{\beta
=1}^{3}a_{\beta }^{4}-3a_{\alpha }^{4}-\left( a_{1}a_{2}a_{3}\right)
^{2}\left( \sum\limits_{\beta =1}^{3}\frac{1}{a_{\beta }^{2}}+\frac{6}{%
a_{\alpha }^{2}}\right) \right] ^{1/2}\right.  \notag \\
&&\left. +2\sum\limits_{\beta =1}^{3}a_{\beta }^{2}-a_{\alpha }^{2}\right\} .
\label{eq21}
\end{eqnarray}

\noindent By substituting (\ref{eq20}) into (\ref{eq5}), and making the
required differentiation and integration with the help of the Dyson's
theorem one can find the following expressions for the corresponding
resonant dielectric permittivity

\begin{eqnarray}
\varepsilon _{3}^{\left( 2\alpha -1\right) } &=&1+\left\{ \frac{%
a_{1}a_{2}a_{3}}{4}\left( \Lambda _{\alpha }^{\prime }+2a_{\alpha
}^{2}\right) \sum\limits_{\beta =1}^{3}\frac{\left( 1+2\delta _{\alpha \beta
}\right) a_{\beta }^{2}I_{\alpha \beta }}{\Lambda _{\alpha }^{\prime
}-a_{\beta }^{2}}\right\} ^{-1},  \notag \\
\varepsilon _{3}^{\left( 2\alpha \right) } &=&1+\left\{ \frac{a_{1}a_{2}a_{3}%
}{4}\left( \Lambda _{\alpha }^{\prime \prime }+2a_{\alpha }^{2}\right)
\sum\limits_{\beta =1}^{3}\frac{\left( 1+2\delta _{\alpha \beta }\right)
a_{\beta }^{2}I_{\alpha \beta }}{\Lambda _{\alpha }^{\prime \prime
}-a_{\beta }^{2}}\right\} ^{-1},  \label{eq22}
\end{eqnarray}

\noindent where $\delta _{\alpha \beta}$ is the Kronecker's delta-symbol
equal to unity at $\alpha = \beta$, and equal to zero at $\alpha \ne \beta$.

One more type of the plasmons for $n=3$ is due to only one function of the
form

\begin{equation}
\phi _{3}^{\left( 7\right) }=x_{1}x_{2}x_{3}.  \label{eq23}
\end{equation}

\noindent By substituting (\ref{eq23}) into (\ref{eq8}), and using the
Dyson's theorem one can find the corresponding values of the permittivity

\begin{equation}
\varepsilon _{3}^{\left( 7\right) }=1-\left\{ \frac{\left(
a_{1}a_{2}a_{3}\right) ^{3}}{2}\left( \sum\limits_{\alpha =1}^{3}a_{\alpha
}^{-2}\right) I_{123}\right\} ^{-1},  \label{eq24}
\end{equation}

\noindent where $I_{123}=\int\limits_{0}^{\infty }\frac{du}{R^{3}\left(
u\right) }$.

The plasmon electric potentials outside ellipsoid can be found from (\ref%
{eq10})by using (\ref{eq20})-(\ref{eq24}) (see Appendix B).

\begin{center}
\textbf{The case $n=4$ ``hexadecapole'' modes}
\end{center}

Now, there are nine different functions of the plasmon oscillation
potentials. Six of them are as follows ($\alpha =1,2,3$)

\begin{eqnarray}
\phi _{4}^{\left( 2\alpha -1\right) } &=&\frac{x_{1}x_{2}x_{3}}{x_{\alpha }}%
\left( \sum\limits_{\beta =1}^{3}\frac{x_{\beta }^{2}}{\Omega _{\alpha
}^{\prime }-a_{\beta }^{2}}+1\right) ,  \notag \\
\phi _{4}^{\left( 2\alpha \right) } &=&\frac{x_{1}x_{2}x_{3}}{x_{\alpha }}%
\left( \sum\limits_{\beta =1}^{3}\frac{x_{\beta }^{2}}{\Omega _{\alpha
}^{\prime \prime }-a_{\beta }^{2}}+1\right) ,  \label{eq25}
\end{eqnarray}

\noindent where $\Omega _{\alpha }^{\prime }$ and $\Omega _{\alpha }^{\prime
\prime }$ are the roots of equation $\sum\limits_{\beta =1}^{3}\frac{%
3-2\delta _{\alpha \beta }}{\Omega _{\alpha }-a_{\beta }^{2}}=0$:

\begin{eqnarray}
\left\{
\begin{array}{c}
\Omega _{\alpha }^{\prime } \\
\Omega _{\alpha }^{\prime \prime }%
\end{array}%
\right\} &=&\frac{1}{7}\left( 2\sum\limits_{\beta =1}^{3}a_{\beta
}^{2}+a_{\alpha }^{2}\right)  \notag \\
&&\pm \frac{1}{7}\left\{ 4\left( \sum\limits_{\beta =1}^{3}a_{\beta
}^{4}-a_{\alpha }^{4}\right) -9a_{\alpha }^{2}\left( \sum\limits_{\beta
=1}^{3}a_{\beta }^{2}-2a_{\alpha }^{2}\right) +\frac{%
a_{1}^{2}a_{2}^{2}a_{3}^{2}}{a_{\alpha }^{2}}\right\} ^{1/2}.  \label{eq26}
\end{eqnarray}

The rest of the plasmon eigenfunctions are ($\alpha =1,2,3$)

\begin{equation}
\phi _{4}^{\left( 6+\alpha \right) }=\left( \sum\limits_{\beta =1}^{3}\frac{%
x_{\beta }^{2}}{\Xi _{\alpha }^{\prime }-a_{\beta }^{2}}+1\right) \left(
\sum\limits_{\beta =1}^{3}\frac{x_{\beta }^{2}}{\Xi _{\alpha }^{\prime
\prime }-a_{\beta }^{2}}+1\right) ,  \label{eq27}
\end{equation}

\noindent where $\Xi _{\alpha }^{\prime }$ and $\Xi _{\alpha }^{\prime
\prime }$ are three different pairs of the roots of the system of equations:

\begin{eqnarray}
\sum\limits_{\alpha =1}^{3}\frac{1}{\Xi ^{\prime }-a_{\alpha }^{2}}+\frac{4}{%
\Xi ^{\prime }-\Xi ^{\prime \prime }} &=&0,  \notag \\
\sum\limits_{\alpha =1}^{3}\frac{1}{\Xi ^{\prime \prime }-a_{\alpha }^{2}}+%
\frac{4}{\Xi ^{\prime \prime }-\Xi ^{\prime }} &=&0.  \label{eq28}
\end{eqnarray}

\noindent The expressions for $\Xi _{\alpha }^{\prime }$ and $\Xi _{\alpha
}^{\prime \prime }$ are rather complicated, so we do not quote them in the
explicit form.

The corresponding to (\ref{eq25}) values of the permittivity are represented
as

\begin{eqnarray}
\varepsilon _{4}^{\left( 1\right) } &=&1+\frac{4}{a_{1}a_{2}a_{3}}\left\{
\left( a_{2}^{2}+a_{3}^{2}\right) ^{2}\sum\limits_{\alpha =1}^{3}\frac{%
\left( 1+2\delta _{\alpha 2}+2\delta _{\alpha 3}\right) a_{\alpha
}^{2}I_{\alpha 23}}{\Omega _{1}^{\prime }-a_{\alpha }^{2}}\right.  \notag \\
&&\left. -\sum\limits_{\alpha =1}^{3}\frac{\left( 1+2\delta _{\alpha
2}+2\delta _{\alpha 3}\right) a_{\alpha }^{2}\left( a_{3}^{2}I_{\alpha
2}+a_{2}^{2}I_{\alpha 3}\right) }{\Omega _{1}^{\prime }-a_{\alpha }^{2}}%
-2\left( a_{2}^{2}+a_{3}^{2}\right) I_{23}\right\} ^{-1},  \notag \\
\varepsilon _{4}^{\left( 3\right) } &=&1+\frac{4}{a_{1}a_{2}a_{3}}\left\{
\left( a_{3}^{2}+a_{1}^{2}\right) ^{2}\sum\limits_{\alpha =1}^{3}\frac{%
\left( 1+2\delta _{\alpha 3}+2\delta _{\alpha 1}\right) a_{\alpha
}^{2}I_{\alpha 31}}{\Omega _{2}^{\prime }-a_{\alpha }^{2}}\right.  \notag \\
&&\left. -\sum\limits_{\alpha =1}^{3}\frac{\left( 1+2\delta _{\alpha
3}+2\delta _{\alpha 1}\right) a_{\alpha }^{2}\left( a_{1}^{2}I_{\alpha
3}+a_{3}^{2}I_{\alpha 1}\right) }{\Omega _{2}^{\prime }-a_{\alpha }^{2}}%
-2\left( a_{3}^{2}+a_{1}^{2}\right) I_{31}\right\} ^{-1},  \notag \\
\varepsilon _{4}^{\left( 5\right) } &=&1+\frac{4}{a_{1}a_{2}a_{3}}\left\{
\left( a_{1}^{2}+a_{2}^{2}\right) ^{2}\sum\limits_{\alpha =1}^{3}\frac{%
\left( 1+2\delta _{\alpha 1}+2\delta _{\alpha 2}\right) a_{\alpha
}^{2}I_{\alpha 12}}{\Omega _{3}^{\prime }-a_{\alpha }^{2}}\right.  \notag \\
&&\left. -\sum\limits_{\alpha =1}^{3}\frac{\left( 1+2\delta _{\alpha
1}+2\delta _{\alpha 2}\right) a_{\alpha }^{2}\left( a_{2}^{2}I_{\alpha
1}+a_{1}^{2}I_{\alpha 2}\right) }{\Omega _{3}^{\prime }-a_{\alpha }^{2}}%
-2\left( a_{1}^{2}+a_{2}^{2}\right) I_{12}\right\} ^{-1},  \label{eq29}
\end{eqnarray}

\noindent where $I_{\alpha \beta \gamma }=\int\limits_{0}^{\infty }\frac{du}{%
\left( u+a_{\alpha }^{2}\right) \left( u+a_{\beta }^{2}\right) \left(
u+a_{\gamma }^{2}\right) R\left( u\right) }$ ($a,\beta ,\gamma =1,2,3$).
Expressions for $\varepsilon _{4}^{\left( 2\right) }$, $\varepsilon
_{4}^{\left( 4\right) }$, $\varepsilon _{4}^{\left( 6\right) }$ can be
obtained via a substitution $\Omega _{\alpha }^{\prime }\rightarrow \Omega
_{\alpha }^{\prime \prime }$ from the expressions for $\varepsilon
_{4}^{\left( 1\right) }$, $\varepsilon _{4}^{\left( 3\right) }$, $%
\varepsilon _{4}^{\left( 5\right) }$, respectively.

The remaining three values of the plasmon oscillation permittivity that
correspond to functions (\ref{eq27}), are of the form ($\alpha =1,2,3)$

\begin{eqnarray}
\varepsilon _{4}^{\left( 6+\alpha \right) } &=&1+\frac{8}{a_{1}a_{2}a_{3}}%
\left\{ -\sum\limits_{\beta =1}^{3}\sum\limits_{\gamma =1}^{3}\frac{\left(
1+2\delta _{\beta \gamma }\right) a_{\beta }^{2}a_{\gamma }^{2}\left(
a_{\beta }^{2}+a_{\gamma }^{2}\right) I_{\beta \gamma }}{\left( \Xi _{\alpha
}^{\prime }-a_{\beta }^{2}\right) \left( \Xi _{\alpha }^{\prime \prime
}-a_{\gamma }^{2}\right) }\right.  \notag \\
&&\left. +\sum\limits_{\beta =1}^{3}\sum\limits_{\gamma =1}^{3}\frac{\left(
1-2\delta _{\beta \gamma }\right) a_{\beta }^{2}a_{\gamma }^{2}\left(
I_{\beta }+I_{\gamma }\right) }{\left( \Xi _{\alpha }^{\prime }-a_{\beta
}^{2}\right) \left( \Xi _{\alpha }^{\prime \prime }-a_{\gamma }^{2}\right) }%
\right.  \notag \\
&&\left. +4\left( \Xi _{\alpha }^{\prime }+\Xi _{\alpha }^{\prime \prime
}\right) \sum\limits_{\beta =1}^{3}\frac{a_{\beta }^{2}I_{\beta }}{\left(
\Xi _{\alpha }^{\prime }-a_{\beta }^{2}\right) \left( \Xi _{\alpha }^{\prime
\prime }-a_{\beta }^{2}\right) }\right\} ^{-1}.  \label{eq30}
\end{eqnarray}

Expressions for the electric potential outside a nanoellipsoid for $n=4$ can
be found from (\ref{eq10}) by using (\ref{eq25})-(\ref{eq30}).

In the case of the higher-order plasmonic modes ( $n>4)$ the plasmon
potentials and the resonant permittivity values of a triaxial nanoellipsoid
can be found in analogous way.

To conclude this section, we repeat a general approach to solving a problem
of plasmon oscillations in the triaxial nanoellipsoid. First, we find
polynomial solutions of the Laplace equation (\ref{eq6}) (the Niven's
functions of the respective order). This can always be done. The only
problem is to find roots of polynomials of several variables. Then using the
Dyson's theorem we find the integral in Eq.(\ref{eq8}) for the eigenvalue.
As the integral is known to be proportional to the Niven's function it is
sufficient to find only the highest terms over coordinates, or even one of
these terms. The calculation of other terms can be used to check
calculations only. Then, by finding the ratio of one the highest terms of
Niven's function to the analogous highest term of the integral we obtain the
desired value of the resonant permittivity. Using a dispersion law for a
particular substance one can find plasmon frequencies of the modes under
consideration.

\section{Discussion and analysis of the results obtained}

The explicit expressions for plasmon potentials and eigenvalues of the
permittivity obtained in the previous section are of great practical
importance. From a general solution for a particle in arbitrary external
field (\ref{eq4}) it is seen that the presence of plasmonic nanoparticles is
manifested in two ways. On the one hand, the \textit{spatial field
distribution} near nanoparticles is defined by the eigenfunctions of a
corresponding resonant plasmon.So, it is especially important to know
\textit{spatial properties} of the plasmon potentials in the region near
nanoparticles (amplification of local fields for SERS and analogous
applications), and far from nanoparticles (to couple the near- and
far-fields, for the nanoantennas and nanolenses).

The far-field plasmons are characterized by usual multipole moments, so the
ALL multipole moments of ALL plasmonic modes should be known. They may be
found from expansion of plasmon potentials far from a particle (the mode
number is omitted)

\begin{equation}
g\left( r\rightarrow \infty \right) =\frac{1}{r^{3}}\sum\limits_{\alpha
=1}^{3}d_{\alpha }x_{\alpha }+\frac{1}{2r^{5}}\sum\limits_{\alpha ,\beta
=1}^{3}Q_{\alpha \beta }x_{\alpha }x_{\beta }+\frac{1}{6r^{7}}%
\sum\limits_{\alpha ,\beta ,\gamma =1}^{3}O_{\alpha \beta \gamma }x_{\alpha
}x_{\beta }x_{\gamma }+...  \label{eq31}
\end{equation}

\noindent where $\;d_{\alpha },\;Q_{\alpha \beta }$ and $O_{\alpha \beta
\gamma }$ are the dipole, quadrupole, and octupole moments, respectively.
Multipole moments of different plasmon modes may be found easily by using a
operator relation between the interior and exterior Niven's functions \cite%
{ref57,ref59}

\begin{equation}
g_{n}=H_{n}\left( \frac{\partial }{\partial x_{1}},\frac{\partial }{\partial
x_{2}},\frac{\partial }{\partial x_{3}}\right) \left\{ 1+\frac{D^{2}}{%
2\left( 2n+3\right) }+\frac{D^{4}}{8\left( 2n+3\right) \left( 2n+5\right) }%
+...\right\} \frac{1}{r},  \label{eq32}
\end{equation}

\noindent where $H_{n}\left( x_{1},x_{2},x_{3}\right) $ is the polynomial
derived from the interior Niven's functions using the highest coordinate
powers only, and where $D^{2}=\sum\limits_{\alpha =1}^{3}a_{\alpha }^{2}%
\frac{\partial ^{2}}{\partial x_{\alpha }^{2}}$ .

At large distance from the origin, every differentiation increases the
singularity of  $1/r$. The $2^{n}$-th multipole moment of the $n$-th mode
will, therefore, be determined by the principal term of Eq.(\ref{eq32})

\begin{equation}
H_{n}\left( \frac{\partial }{\partial x_{1}},\frac{\partial }{\partial x_{2}}%
,\frac{\partial }{\partial x_{3}}\right) \frac{1}{r}=\frac{\left( -1\right)
^{n}\left( 2n\right) !}{2^{n}n!}\frac{1}{r^{2n+1}}H_{n}\left(
x_{1},x_{2},x_{3}\right) .  \label{eq33}
\end{equation}

\noindent By comparing that expression with the expansion
(\ref{eq31}) one may find the desired $2^{n}$-th multipole moments
of $n$-th mode. The multipole moments of higher powers ($2^{n+1}$,
$2^{n+2}$...) can be found from (\ref{eq32}) by keeping in series
the next terms $D^{2}$, $D^{4}$, \ldots The multipole moments of
lower powers ($2^{n-1}$, $2^{n-2}$...) for $n$-th mode will be equal
to zero.

On the other hand, the \textit{amplitude} of plasmon resonance is defined by
efficiency of a plasmon excitation. In its turn this efficiency is product
of 2 factors. First is the integral of spatial overlapping of external and
plasmon field (factor $\int\limits_{V}d\mathbf{r}\left( \mathbf{e}_{n}\cdot
\mathbf{E}^{0}\right) $ in Eq.(\ref{eq4})). Second factor is determined by a
coincidence of external field frequencies with the resonant plasmon
frequency (factor $\left( \varepsilon _{n}-\varepsilon \left( \omega \right)
\right) ^{-1}$ in (\ref{eq4})). The last one depends considerably on
properties of a nanoparticle material. The described features of plasmon
resonances in nanoparticles will be analyzed below.

\begin{center}
\textbf{The case $n=1$ ``dipole'' modes}
\end{center}

The electric field of "dipole" modes inside an ellipsoid nanoparticle is
homogeneous (\ref{eq11}), and these modes will be most effectively excited
by the homogeneous external fields which maximaze overlapping integral $%
\int\limits_{V}d\mathbf{r}\left( \mathbf{e}_{n}\cdot \mathbf{E}^{0}\right) $%
. In the case of a highly nonhomogeneous external field, the spacial
variations of the field will be averaged and equivalent to a weaker
homogeneous field.

A spatial distribution of the squared electric field for modes with $n=1$ is
shown in Fig.2 where the ellipsoid axes are $a_{2}/a_{1}=0.6$ and $%
a_{3}/a_{1}=0.4$. It is seen that for a given geometry there are always two
maxima localized at a nanoparticle surface. The mode whose field is along
the longest axis has the highest nonhomogeneity (Fig.2, $m=1$). The maximal
amplification of the external field by that mode is determined by factor

\begin{equation}
\left\vert \frac{E}{E^{0}}\right\vert =\left\vert \frac{\left( \varepsilon
_{1}^{\left( 1\right) }-1\right) \varepsilon \left( \omega \right) }{%
\varepsilon _{1}^{\left( 1\right) }-\varepsilon \left( \omega \right) }%
\right\vert .  \label{eq34}
\end{equation}

Figure 3 illustrates the surface charge distribution for modes with $n=1$.
One can see a quasidipole character of those modes. But unlike the spherical
plasmons, the plasmons of ellipsoid nanoparticles have the high-order
multipole moments. For the case $n=1$, the quadrupole moment of the mode
proves to be zero. For dipole and octupole moments we obtain (see Eq.(\ref%
{eq31})).

A) For the mode $g_{1}^{\left( 1\right) }$ (for more information see
Appendix B)

\begin{eqnarray}
d_{\alpha } &=&-\delta _{\alpha 1}\frac{1}{4\pi }\left( \varepsilon
_{1}^{\left( 1\right) }-1\right) V,  \notag \\
O_{\alpha \beta \gamma } &=&\delta _{\alpha \beta }\delta _{\gamma 1}\frac{9%
}{20\pi }\left( \varepsilon _{1}^{\left( 1\right) }-1\right) \left\{
2a_{1}^{2}-5a_{\alpha }^{2}+\sum\limits_{\sigma =1}^{3}a_{\sigma
}^{2}\right\} V.  \label{eq35}
\end{eqnarray}

B) For the mode $g_{1}^{\left( 2\right) }$

\begin{eqnarray}
d_{\alpha } &=&-\delta _{\alpha 2}\frac{1}{4\pi }\left( \varepsilon
_{1}^{\left( 2\right) }-1\right) V,  \notag \\
O_{\alpha \beta \gamma } &=&\delta _{\alpha \beta }\delta _{\gamma 2}\frac{9%
}{20\pi }\left( \varepsilon _{1}^{\left( 2\right) }-1\right) \left\{
2a_{2}^{2}-5a_{\alpha }^{2}+\sum\limits_{\sigma =1}^{3}a_{\sigma
}^{2}\right\} V.  \label{eq36}
\end{eqnarray}

C) For the mode $g_{1}^{\left( 3\right) }$

\begin{eqnarray}
d_{\alpha } &=&-\delta _{\alpha 3}\frac{1}{4\pi }\left( \varepsilon
_{1}^{\left( 3\right) }-1\right) V,  \notag \\
O_{\alpha \beta \gamma } &=&\delta _{\alpha \beta }\delta _{\gamma 3}\frac{9%
}{20\pi }\left( \varepsilon _{1}^{\left( 3\right) }-1\right) \left\{
2a_{3}^{2}-5a_{\alpha }^{2}+\sum\limits_{\sigma =1}^{3}a_{\sigma
}^{2}\right\} V.  \label{eq37}
\end{eqnarray}

In the case of a sphere ($a_{1} = a_{2} = a_{3} $), the dipole mode has a
zero octupole moment, as follows from (\ref{eq35})-(\ref{eq37}).

The resonant values of dielectric constant are connected with a
corresponding plasmon frequency. Above we found the values of resonance
permittivity that are independent of a nanoparticle material. To calculate
the plasmon frequencies it is necessary to use the corresponding\ dispersion
laws for real substances. Figure 4 demonstrates the plasmon frequency
dependences of the modes with $n=1$ for the Drude's dispersion law $\omega
_{n}=\omega _{pl}{/}\sqrt{1-\varepsilon _{n}}$. It is seen that the plasmon
frequencies of all modes have a simple monotonic character. As a whole, the
properties of modes with $n=1$ are quite transparent and analogous to dipole
modes of a sphere.

\begin{center}
\textbf{The case $n=2$ ``qudrupole'' modes}
\end{center}

Unlike the case with $n=1$, the modes with $n=2$ may be of two types. They
are even modes (relative to  change of sign of one of the coordinates, two
modes) (\ref{eq15}) , and odd modes (\ref{eq18}) (three modes). Such modes
cannot be excited by a homogeneous field of any frequency, because of a
overlap integral is equal to zero for such modes. On the other hand, such
modes may be excited by fields that are highly nonhomogeneous in the
vicinity of a nanoparticle.

A spatial distribution of a squared electric field of the modes with $n=2$
is shown in Fig.5 for an ellipsoid with axes $a_{2}/a_{1}=0.6$ and $%
a_{3}/a_{1}=0.4$. From this Figure it is seen that for a given geometry
there are four maxima and one local minimum of the field. Figure 6
illustrates the surface charge distribution for the same modes. It shows
vividly a quasiquadrupole character of these modes.

Below we draw down explicit equations for the dipole, quadrupole, and
octopole moments of the modes with $n=2$ (see Eq.(\ref{eq31})). The dipole
and octopole moments are zero in this case. For a quadrupole moment we have
the following expressions:

A) For the mode $g_{2}^{\left( 1\right) }$ (for more information see
Appendix B)

\begin{equation}
Q_{\alpha \beta }=-\delta _{\alpha \beta }\frac{3}{5\pi \left( \Lambda
^{\prime }-a_{\alpha }^{2}\right) }\left( \varepsilon _{2}^{\left( 1\right)
}-1\right) \left( \sum\limits_{\gamma =1}^{3}a_{\gamma }^{2}\right) V.
\label{eq38}
\end{equation}

B) For the mode $g_{2}^{\left( 2\right) }$

\begin{equation}
Q_{\alpha \beta }=-\delta _{\alpha \beta }\frac{3}{5\pi \left( \Lambda
^{\prime \prime }-a_{\alpha }^{2}\right) }\left( \varepsilon _{2}^{\left(
2\right) }-1\right) \left( \sum\limits_{\gamma =1}^{3}a_{\gamma }^{2}\right)
V.  \label{eq39}
\end{equation}

C) For the mode $g_{2}^{\left( 3\right) }$

\begin{equation}
Q_{\alpha \beta }=-\frac{3}{10\pi }\left( \varepsilon _{2}^{\left( 3\right)
}-1\right) \left( \delta _{\alpha 2}\delta _{\beta 3}a_{2}^{2}+\delta
_{\alpha 3}\delta _{\beta 2}a_{3}^{2}\right) V.  \label{eq40}
\end{equation}

D) For the mode $g_{2}^{\left( 4\right) }$

\begin{equation}
Q_{\alpha \beta }=-\frac{3}{10\pi }\left( \varepsilon _{2}^{\left( 4\right)
}-1\right) \left( \delta _{\alpha 1}\delta _{\beta 3}a_{1}^{2}+\delta
_{\alpha 3}\delta _{\beta 1}a_{3}^{2}\right) V.  \label{eq41}
\end{equation}

E) For the mode $g_{2}^{\left( 5\right) }$

\begin{equation}
Q_{\alpha \beta }=-\frac{3}{10\pi }\left( \varepsilon _{2}^{\left( 5\right)
}-1\right) \left( \delta _{\alpha 1}\delta _{\beta 2}a_{1}^{2}+\delta
_{\alpha 2}\delta _{\beta 1}a_{2}^{2}\right) V.  \label{eq42}
\end{equation}

Figure 7 demonstrates the dependences of plasmon frequencies of the modes
with $n=2$ for the Drude's dispersion law on semi-axis ratio. The plasmon
frequencies of the modes with $n=2$ have a nontrivial character with local
maxima and minima. In the case of nanoellipsoids of arbitrary shape, the
surface $\omega _{2}^{\left( m\right) }\left( a_{3}/a_{1},a_{2}/a_{1}\right)
$ has the form of valleys and ridges and has no extremums for nontrivial
ellipsoids. From quantum condition that the plasmon frequency is
proportional to the plasmon energy it becomes clear, from thermodynamic
viewpoint, that nanoparticles with an excited mode are under strain that is
minimal at certain geometries. This circumstance could be used in a
synthesis of nanoparticles which are optimal for the given plasmons.

From the analysis of Fig.7 it also follows that under certain ratios of $%
a_{3}/a_{1}$ ($a_{2}/a_{1}$ is fixed), the values of plasmon frequencies
corresponding to different indices $m=1,2,\ldots ,5$ of the multipolarity $%
n=2$, may coincide. So, by using a nonhomogeneous source of electromagnetic
field one may simultaneously excite several different types (modes) of the
plasmon oscillations at one frequency. This fact can be used for development
of efficient souces of light and related devices\cite{ref50} . When
ellipsoid is degenerated into a spheroid ($a_{1}=a_{2}\neq a_{3}$), some of
the curves represented in Fig.7 should coincide. In case of multipolarity $%
n=2$, there should remain only $n+1=3$ independents dispersion curves (and
not $2n+1=5$, as for an ellipsoid of general shape). In a particular case of
a sphere ($a_{1}=a_{2}=a_{3}$) there will remain only one dispersion point
to be determined by the well known expression $\omega =\omega _{pl}\sqrt{2/5}
$.

As a whole, the modes with $n=2$, though having much in common with
quadrupole modes in a sphere, have a variety of nontrivial properties and
may be successfully applied in different nanooptical devices.

\begin{center}
\textbf{The case $n=3$ ``octupole'' modes}
\end{center}

The modes with $n=3$ are described by Eqs. (\ref{eq22}) (six modes) and (\ref%
{eq23}) (one mode). They can not be excited by a homogeneous field,
because one can show that their dipole moments are equal to zero.
However, such modes may be excited by fields which are
nonhomogeneous in the vicinity of a particle.

The squared electric field distribution of these modes is shown in Fig.8. It
is seen that the field structure is very complex and has several maxima and
minima. Note also that the squared field maxima are always on a surface of a
nanoparticle.

A surface charge distribution of modes with $n=3$, as shown in Fig.9, is
rather complex too. Nevertheless a spatial field structure of those modes
can be clearly understood from a simultaneous analysis of Figs. 8 and 9.

Figure 10 illustrates dependences of the plasmon frequencies on one of the
ellipsoid semi-axes. These dependences also have the maxima and the minima
that can be used in a number of spectroscopic applications.

\begin{center}
\textbf{The case $n=4$ ``hexadecapole'' modes}
\end{center}

Spatial structure of the modes with $n=4$ is more complex, and we do not
consider it in this work. We restrict ourselves to dependences of the
relative plasmon frequencies $\omega _{4}^{\left( m\right) }/\omega _{pl}$ ($%
m=1,2,...,9)$ on one of the semi-axes $a_{3}/a_{1}$ ration at the given
ratio $a_{2}/a_{1}=0.6$. (Fig.11). These dependences are similar, as a
whole, to the lower-order mode dependences, but now one dipsersion curve may
have several maxima and minima.

\begin{center}
\textbf{The case $n>4$}
\end{center}

As it was mentioned already higher orders plasmonic modes in 3-axial
nanoellipsoid can be found analytically within our general approach.
Such analytical solution is very complicated  and we have restricted
our analytical calculations to modes with $n\leq 4$ here.

Nevertheless for the sake of completeness it is important to know
(at least qualitatively) properties of all higher modes. To
calculate higher
orders modes we made use of boundary elements method (BEM, see e.g. \cite%
{ref64}). This method is based on surface integral equation

\begin{equation}
2\pi \sigma \left( Q\right) =\frac{\varepsilon _{n}-1}{\varepsilon _{n}+1}%
\int_{S} dS_{M} \sigma \left( M\right) \frac{ \left( \mathbf{r}_{MQ}\cdot\mathbf{n}_{Q} \right) }{%
r_{MQ}^{3}},  \label{eq43a}
\end{equation}

\noindent which can be derived from volume integral equation
(\ref{eq5}). Here $\sigma \left( Q\right)$ is surface charge density
\ of plasmon mode, $\mathbf{n}_{Q}$ is normal to surface of
nanoparticles at point $Q$ , $\mathbf{r}_{MQ}$ is radius-vector from
point $M$ to point $Q$ and integration is over surface of
nanoparticle.Then one can partition surface $S$ of nanoparticle into
$N$ small triangle pieces $\Delta S$ and reduce surface integral
equation to system of $N$ linear equations.

The results of numerical calculations  with 3000 triangles are shown
in Fig.12, where dispersion curves of  300 modes are shown  by
circles while solid lines are analytical dispersion curves from
Figures 4, 7 and 10.

From this Figure the general structure of dispersion curves of
plasmonic modes in triaxial nanoellipsoid is clearly seen. In the
region $a_{3}/a_{1}>0.1$ there is a good agreement between
analytical and numerical calculations. On the contrary, in the
region $a_{3}/a_{1}<0.1$ numerical solution is inaccurate and \
differs substantially from analytical one because of very sharp
edges of ellipsoid in this region. The analytical solution show that
resonat permittivity goes to infinity or to zero in the limit of
very thin ellipsoid $\left( a_{3}/a_{1}\rightarrow 0\right)$.

\section{Conclusion}

In this work, the plasmon oscillations of a triaxial nanoellipsoid
have been studied in a quasistatic approximation. A general approach
has been proposed to finding analytical expressions for the plasmon
electric potentials and the corresponding values of permittivity (or
plasmon frequencies) of an arbitrary multipolarity order. By making
use of that approach one could derive the explicit analytical
expressions for potentials of several plasmon modes and the
dielectric permittivity.

The analytical expressions obtained in this work are extremely important for
nanoplasmonics and nanooptics. They can be used to interpret the
experimental results on optical properties of nanoparticles and to assess
reliability of numerical calculations on the interaction of electromagnetic
radiation with nanoparticles that may be approximated by triaxial
nanoellipsoids.

The variety and complexity of optical properties of the triaxial
nanoellipsoids can also be used in many new nanooptical applications. In
particular, by choosing a suitable geometry of an ellipsoid one can ensure
its effective interaction with elementary quantum emitters both in the
radiation and absorption regimes. As a result, the efficiency of natural and
artificial fluorophores and other nanodimensional light sources might be
substantially enhanced \cite{ref50,ref62}.

In addition, due to the complexity of geometry of an ellipsoid and
its plasmon modes, one might build, on its basis, the effective
optical nanoantennas \cite{ref15}-\cite{ref18}. They possess all
properties of more complex in production Yagi-Uda nanoantennas
\cite{ref63}, and may be more easily synthesized. Of course, a
detailed description of a non-dipole radiation pattern of an
ellipsoid nanoantenna is possible only beyond the framework of
quasistatic approximation only. However, the results obtained in
this work make it possible to perform fast estimations. Moreover,
within the framework of the longwavelength perturbation theory \ one
can find next order corrections ($\sim 1/\lambda ^{2}$ and $\sim
1/\lambda ^{3}$ ) to resonant values of permittivity and the
respective eigenfunctions \cite{ref62}.

\begin{acknowledgments}
D.V.G. is grateful to the Belarusian Republican Fund for Fundamental
Research (grant F08M-080) for financial support of the present work.

V.V.K. and M.P. are grateful to Russian Foundation for Basic
Researches (grants 07-02-01328 and 05-02-19647), the RAS Presidium
Program ``Quantum Macrophysics'', and ``Advanced Energy
Technologies'', LTD, for partial financial support of the present
work.
\end{acknowledgments}

\appendix

\section{The explicit expressions for the integrals used in
calculations of the plasmon potentials of a triaxial nanoellipsoid}

By F.D. Dyson the following theorem has been proved \cite{ref60,ref61} ($%
p,q,s=0,1,2,\ldots $)

\begin{eqnarray}
\int\limits_{V}d\mathbf{r}^{\prime }\frac{x_{1}^{\prime }{}^{p}x_{2}^{\prime
}{}^{q}x_{3}^{\prime }{}^{s}}{\left\vert \mathbf{r}^{\prime }-\mathbf{r}%
\right\vert } &=&\sum\limits_{k=0}^{\left[ N/2\right] }\frac{\pi
a_{1}a_{2}a_{3}}{2^{2k}k!\left( k+1\right) !}  \notag \\
&&\times \int\limits_{u_{0}}^{\infty }\frac{duu^{k}}{R\left( u\right) }%
\left( 1-\sum\limits_{\alpha =1}^{3}\frac{x_{\alpha }^{2}}{u+a_{\alpha }^{2}}%
\right) ^{k+1}  \notag \\
&&\times \hat{D}^{k}\left\{ \left( \frac{a_{1}^{2}x_{1}}{u+a_{1}^{2}}\right)
^{p}\left( \frac{a_{2}^{2}x_{2}}{u+a_{2}^{2}}\right) ^{q}\left( \frac{%
a_{3}^{2}x_{3}}{u+a_{3}^{2}}\right) ^{s}\right\} ,  \label{eq43}
\end{eqnarray}

\noindent where $N$=max($p,q,s$); $\hat{D}=\sum\limits_{\alpha =1}^{3}\frac{%
u+a_{\alpha }^{2}}{a_{\alpha }^{2}}\frac{\partial ^{2}}{\partial x_{\alpha
}^{2}}$, $\hat{D}^{0}=1$. Basing on this theorem one can obtain analytical
expressions for the following integrals used in this work ($\alpha ,\beta
,\gamma ,\sigma =1,2,3$):

\begin{equation}
\int\limits_{V}\frac{d\mathbf{r}^{\prime }}{\left\vert \mathbf{r}^{\prime }-%
\mathbf{r}\right\vert }=\pi a_{1}a_{2}a_{3}\left( I\left( u_{0}\right)
-\sum\limits_{\alpha =1}^{3}x_{\alpha }^{2}I_{\alpha }\left( u_{0}\right)
\right) ,  \label{eq44}
\end{equation}

\begin{equation}
\int\limits_{V}d\mathbf{r}^{\prime }\frac{x_{\alpha }^{\prime }}{\left\vert
\mathbf{r}^{\prime }-\mathbf{r}\right\vert }=\pi a_{1}a_{2}a_{3}a_{\alpha
}^{2}x_{\alpha }\left( I_{\alpha }\left( u_{0}\right) -\sum\limits_{\beta
=1}^{3}x_{\beta }^{2}I_{\alpha \beta }\left( u_{0}\right) \right) ,
\label{eq45}
\end{equation}

\begin{eqnarray}
\int\limits_{V}d\mathbf{r}^{\prime }\frac{x_{\alpha }^{\prime }x_{\beta
}^{\prime }}{\left\vert \mathbf{r}^{\prime }-\mathbf{r}\right\vert } &=&\pi
a_{1}a_{2}a_{3}a_{\alpha }^{2}a_{\beta }^{2}x_{\alpha }x_{\beta }\left(
I_{\alpha \beta }\left( u_{0}\right) -\sum\limits_{\gamma =1}^{3}x_{\gamma
}^{2}I_{\alpha \beta \gamma }\left( u_{0}\right) \right)  \notag \\
&&+\frac{1}{4}\pi a_{1}a_{2}a_{3}a_{\alpha }^{2}\delta _{\alpha \beta
}\left( I\left( u_{0}\right) -2\sum\limits_{\gamma =1}^{3}x_{\gamma
}^{2}I_{\gamma }\left( u_{0}\right) \right.  \notag \\
&&\left. +\sum\limits_{\gamma =1}^{3}\sum\limits_{\sigma =1}^{3}x_{\gamma
}^{2}x_{\sigma }^{2}I_{\gamma \sigma }\left( u_{0}\right) \right)  \notag \\
&&-\frac{1}{4}\pi a_{1}a_{2}a_{3}a_{\alpha }^{4}\delta _{\alpha \beta
}\left( I_{\alpha }\left( u_{0}\right) -2\sum\limits_{\gamma
=1}^{3}x_{\gamma }^{2}I_{\alpha \gamma }\left( u_{0}\right) \right.  \notag
\\
&&\left. +\sum\limits_{\gamma =1}^{3}\sum\limits_{\sigma =1}^{3}x_{\gamma
}^{2}x_{\sigma }^{2}I_{\alpha \gamma \sigma }\left( u_{0}\right) \right) ,
\label{eq46}
\end{eqnarray}

\begin{eqnarray}
\int\limits_{V}d\mathbf{r}^{\prime }\frac{x_{\alpha }^{\prime }x_{\beta
}^{\prime }x_{\gamma }^{\prime }}{\left\vert \mathbf{r}^{\prime }-\mathbf{r}%
\right\vert } &=&\pi a_{1}a_{2}a_{3}a_{\alpha }^{2}a_{\beta }^{2}a_{\gamma
}^{2}x_{\alpha }x_{\beta }x_{\gamma }\left( I_{\alpha \beta \gamma }\left(
u_{0}\right) -\sum\limits_{\sigma =1}^{3}x_{\sigma }^{2}I_{\alpha \beta
\gamma \sigma }\left( u_{0}\right) \right)  \notag \\
&&+\frac{1}{4}\pi a_{1}a_{2}a_{3}\delta _{\alpha \beta }a_{\alpha
}^{2}a_{\gamma }^{2}x_{\gamma }\left( I_{\alpha }\left( u_{0}\right)
-2\sum\limits_{\sigma =1}^{3}x_{\sigma }^{2}I_{\alpha \sigma }\left(
u_{0}\right) \right.  \notag \\
&&\left. +\sum\limits_{\sigma =1}^{3}\sum\limits_{\rho =1}^{3}x_{\sigma
}^{2}x_{\rho }^{2}I_{\alpha \sigma \rho }\left( u_{0}\right) \right)  \notag
\\
&&-\frac{1}{4}\pi a_{1}a_{2}a_{3}\delta _{\alpha \beta }a_{\alpha
}^{2}a_{\gamma }^{4}x_{\gamma }\left( I_{\alpha \gamma }\left( u_{0}\right)
-2\sum\limits_{\sigma =1}^{3}x_{\sigma }^{2}I_{\alpha \gamma \sigma }\left(
u_{0}\right) \right.  \notag \\
&&\left. +\sum\limits_{\sigma =1}^{3}\sum\limits_{\sigma =1}^{3}x_{\sigma
}^{2}x_{\rho }^{2}I_{\alpha \gamma \sigma \rho }\left( u_{0}\right) \right)
\notag \\
&&+\frac{1}{2}\pi a_{1}a_{2}a_{3}\delta _{\alpha \beta }\delta _{\alpha
\gamma }a_{\alpha }^{4}x_{\alpha }\left( I_{\alpha }\left( u_{0}\right)
-2\sum\limits_{\sigma =1}^{3}x_{\sigma }^{2}I_{\alpha \sigma }\left(
u_{0}\right) \right.  \notag \\
&&\left. +\sum\limits_{\sigma =1}^{3}\sum\limits_{\rho =1}^{3}x_{\sigma
}^{2}x_{\rho }^{2}I_{\alpha \sigma \rho }\left( u_{0}\right) \right)  \notag
\\
&&-\frac{1}{2}\pi a_{1}a_{2}a_{3}\delta _{\alpha \beta }\delta _{\alpha
\gamma }a_{\alpha }^{6}x_{\alpha }\left( I_{\alpha \alpha }\left(
u_{0}\right) -2\sum\limits_{\sigma =1}^{3}x_{\sigma }^{2}I_{\alpha \alpha
\sigma }\left( u_{0}\right) \right.  \notag \\
&&\left. +\sum\limits_{\sigma =1}^{3}\sum\limits_{\rho =1}^{3}x_{\sigma
}^{2}x_{\rho }^{2}I_{\alpha \alpha \sigma \rho }\left( u_{0}\right) \right) ,
\label{eq47}
\end{eqnarray}

\noindent in which $\delta _{\alpha \beta} $ is the Kronecker's
delta-symbol equal to unity at $\alpha = \beta $, and equal to zero
at $\alpha \ne \beta $;

\begin{eqnarray}
I\left( u_{0}\right) &=&\int\limits_{u_{0}}^{\infty }\frac{du}{R\left(
u\right) },\quad I_{\alpha }\left( u_{0}\right) =\int\limits_{u_{0}}^{\infty
}\frac{du}{\left( u+a_{\alpha }^{2}\right) R\left( u\right) },  \notag \\
I_{\alpha \beta }\left( u_{0}\right) &=&\int\limits_{u_{0}}^{\infty }\frac{du%
}{\left( u+a_{\alpha }^{2}\right) \left( u+a_{\beta }^{2}\right) R\left(
u\right) },  \notag \\
I_{\alpha \beta \gamma }\left( u_{0}\right) &=&\int\limits_{u_{0}}^{\infty }%
\frac{du}{\left( u+a_{\alpha }^{2}\right) \left( u+a_{\beta }^{2}\right)
\left( u+a_{\gamma }^{2}\right) R\left( u\right) },  \notag \\
I_{\alpha \beta \gamma \sigma }\left( u_{0}\right)
&=&\int\limits_{u_{0}}^{\infty }\frac{du}{\left( u+a_{\alpha }^{2}\right)
\left( u+a_{\beta }^{2}\right) \left( u+a_{\gamma }^{2}\right) \left(
u+a_{\sigma }^{2}\right) R\left( u\right) },  \label{eq48}
\end{eqnarray}

\noindent where $R\left( u\right) =\prod\limits_{\alpha =1}^{3}\sqrt{%
u+a_{\alpha }^{2}}$; the parameter $u_{0}=0$, if the point $\mathbf{r}$ is
inside an ellipsoid, and $u_{0}$ is the positive root of equation $%
\sum\limits_{\alpha =1}^{3}\frac{x_{\alpha }^{2}}{u_{0}+a_{\alpha }^{2}}=1$,
if point $\mathbf{r}$ is outside the ellipsoid.

\section{The explicit expressions for plasmon wavefunction outside
ellipsoid (see Eq.(\ref{eq10}))}

A) In the case $n=1$ we get three functions ($\alpha =1,2,3$)

\begin{equation}
g_{1}^{\left( \alpha \right) }=x_{\alpha }\frac{I_{\alpha }\left(
u_{0}\right) }{I_{\alpha }}.  \label{eq49}
\end{equation}

B) In the case $n=2$ we get five functions. Two of them are of the form

\begin{eqnarray}
g_{2}^{\left( 1\right) } &=&\left( \sum\limits_{\beta =1}^{3}\frac{a_{\beta
}^{2}I_{\beta }}{\Lambda ^{\prime }-a_{\beta }^{2}}\right) ^{-1}\left\{
\sum\limits_{\beta =1}^{3}\frac{a_{\beta }^{2}I_{\beta }\left( u_{0}\right)
}{\Lambda ^{\prime }-a_{\beta }^{2}}\left( \sum\limits_{\alpha =1}^{3}\frac{%
x_{\alpha }^{2}}{\Lambda ^{\prime }-a_{\alpha }^{2}}+1\right) \right.  \notag
\\
&&\left. +\frac{2u_{0}}{R\left( u_{0}\right) }\sum\limits_{\alpha =1}^{3}%
\frac{x_{\alpha }^{2}}{\left( \Lambda ^{\prime }-a_{\alpha }^{2}\right)
\left( u_{0}+a_{\alpha }^{2}\right) }\right\} ,  \notag \\
g_{2}^{\left( 2\right) } &=&\left( \sum\limits_{\beta =1}^{3}\frac{a_{\beta
}^{2}I_{\beta }}{\Lambda ^{\prime \prime }-a_{\beta }^{2}}\right)
^{-1}\left\{ \sum\limits_{\beta =1}^{3}\frac{a_{\beta }^{2}I_{\beta }\left(
u_{0}\right) }{\Lambda ^{\prime \prime }-a_{\beta }^{2}}\left(
\sum\limits_{\alpha =1}^{3}\frac{x_{\alpha }^{2}}{\Lambda ^{\prime \prime
}-a_{\alpha }^{2}}+1\right) \right.  \notag \\
&&\left. +\frac{2u_{0}}{R\left( u_{0}\right) }\sum\limits_{\alpha =1}^{3}%
\frac{x_{\alpha }^{2}}{\left( \Lambda ^{\prime \prime }-a_{\alpha
}^{2}\right) \left( u_{0}+a_{\alpha }^{2}\right) }\right\} ,  \label{eq50}
\end{eqnarray}

\noindent where

\begin{equation}
\left\{
\begin{array}{c}
\Lambda ^{\prime } \\
\Lambda ^{\prime \prime }%
\end{array}%
\right\} =\frac{1}{3}\left\{ \pm \left[ \sum\limits_{\alpha =1}^{3}\left(
a_{\alpha }^{4}-\frac{a_{1}^{2}a_{2}^{2}a_{3}^{2}}{a_{\alpha }^{2}}\right) %
\right] ^{1/2}+\sum\limits_{\alpha =1}^{3}a_{\alpha }^{2}\right\} .
\label{eq51}
\end{equation}

The rest three have the form

\begin{equation}
g_{2}^{\left( 3\right) }=x_{2}x_{3}\frac{I_{23}\left( u_{0}\right) }{I_{23}}%
,\quad g_{2}^{\left( 4\right) }=x_{3}x_{1}\frac{I_{31}\left( u_{0}\right) }{%
I_{31}},\quad g_{2}^{\left( 5\right) }=x_{1}x_{2}\frac{I_{12}\left(
u_{0}\right) }{I_{12}}.  \label{eq52}
\end{equation}

C) In the case $n=3$ we get seven functions. Six of them are
($\alpha =1,2,3$)

\begin{eqnarray}
g_{3}^{\left( 2\alpha -1\right) } &=&x_{\alpha }\left( \sum\limits_{\gamma
=1}^{3}\frac{\left( 1+2\delta _{\alpha \gamma }\right) a_{\gamma
}^{2}I_{\alpha \gamma }}{\Lambda _{\alpha }^{\prime }-a_{\gamma }^{2}}%
\right) ^{-1}  \notag \\
&&\times \left\{ \sum\limits_{\gamma =1}^{3}\frac{\left( 1+2\delta _{\alpha
\gamma }\right) a_{\gamma }^{2}I_{\alpha \gamma }\left( u_{0}\right) }{%
\Lambda _{\alpha }^{\prime }-a_{\gamma }^{2}}\left( \sum\limits_{\beta
=1}^{3}\frac{x_{\beta }^{2}}{\Lambda _{\alpha }^{\prime }-a_{\beta }^{2}}%
+1\right) \right.  \notag \\
&&\left. +\frac{2u_{0}}{R\left( u_{0}\right) \left( \Lambda _{\alpha
}^{\prime }+2a_{\alpha }^{2}\right) \left( u_{0}+a_{\alpha }^{2}\right) }%
\left( \sum\limits_{\beta =1}^{3}\frac{\left( 2a_{\alpha }^{2}+a_{\beta
}^{2}\right) x_{\beta }^{2}}{\left( \Lambda _{\alpha }^{\prime }-a_{\beta
}^{2}\right) \left( u_{0}+a_{\beta }^{2}\right) }+1\right) \right\} ,
\label{eq53}
\end{eqnarray}

\noindent and

\begin{eqnarray}
g_{3}^{\left( 2\alpha \right) } &=&x_{\alpha }\left( \sum\limits_{\gamma
=1}^{3}\frac{\left( 1+2\delta _{\alpha \gamma }\right) a_{\gamma
}^{2}I_{\alpha \gamma }}{\Lambda _{\alpha }^{\prime \prime }-a_{\gamma }^{2}}%
\right) ^{-1}  \notag \\
&&\times \left\{ \sum\limits_{\gamma =1}^{3}\frac{\left( 1+2\delta _{\alpha
\gamma }\right) a_{\gamma }^{2}I_{\alpha \gamma }\left( u_{0}\right) }{%
\Lambda _{\alpha }^{\prime \prime }-a_{\gamma }^{2}}\left(
\sum\limits_{\beta =1}^{3}\frac{x_{\beta }^{2}}{\Lambda _{\alpha }^{\prime
\prime }-a_{\beta }^{2}}+1\right) \right.  \notag \\
&&\left. +\frac{2u_{0}}{R\left( u_{0}\right) \left( \Lambda _{\alpha
}^{\prime \prime }+2a_{\alpha }^{2}\right) \left( u_{0}+a_{\alpha
}^{2}\right) }\left( \sum\limits_{\beta =1}^{3}\frac{\left( 2a_{\alpha
}^{2}+a_{\beta }^{2}\right) x_{\beta }^{2}}{\left( \Lambda _{\alpha
}^{\prime \prime }-a_{\beta }^{2}\right) \left( u_{0}+a_{\beta }^{2}\right) }%
+1\right) \right\} ,  \label{eq54}
\end{eqnarray}

\noindent where

\begin{eqnarray}
\left\{
\begin{array}{c}
\Lambda _{\alpha }^{\prime } \\
\Lambda _{\alpha }^{\prime \prime }%
\end{array}%
\right\} &=&\frac{1}{5}\left\{ \pm \left[ 4\sum\limits_{\beta
=1}^{3}a_{\beta }^{4}-3a_{\alpha }^{4}-\left( a_{1}a_{2}a_{3}\right)
^{2}\left( \sum\limits_{\beta =1}^{3}\frac{1}{a_{\beta }^{2}}+\frac{6}{%
a_{\alpha }^{2}}\right) \right] ^{1/2}\right.  \notag \\
&&\left. +2\sum\limits_{\beta =1}^{3}a_{\beta }^{2}-a_{\alpha }^{2}\right\} .
\label{eq55}
\end{eqnarray}

The rest seventh function is

\begin{equation}
g_{3}^{\left( 7\right) }=x_{1}x_{2}x_{3}\frac{I_{123}\left( u_{0}\right) }{%
I_{123}}.  \label{eq56}
\end{equation}

\noindent In the expressions above we have used notations $I_{\alpha
}=I_{\alpha }\left( 0\right) $, $I_{\alpha \beta }=I_{\alpha \beta }\left(
0\right) $ and $I_{\alpha \beta \gamma }=I_{\alpha \beta \gamma }\left(
0\right) $ (see Eq. (\ref{eq48})).

It should be noted, if we set $u_{0} = 0$ (ellipsoid surface), then the
potential functions of an exterior problem, represented in Appendix B, turn
into the respective functions of an interior problem.

\pagebreak

\newpage

\begin{center}
\bigskip {\LARGE List of Figure Captions}
\end{center}

\bigskip

Fig.1 Geometry of the problem.

Fig.2 (Color online) Distribution of the squared electric field logarithm
(a.u.) near a triaxial nanoellipsoid with semi-axes $a_{2}/a_{1}=0.6$ and $%
a_{3}/a_{1}=0.4$ in the plane $x_{3}=0$ for the plasmon modes with $n=1$.
Index $m=1,2,3$ denotes upper index of the potential function $\phi
_{1}^{\left( m\right) }$. Transition from red color to blue corresponds to
transition from the highest squared values of electric field to the least
squared values. Black line corresponds to the surface of the nanoparticle.

Fig.3 (Color online) Surface charge distribution (a.u.) of plasmon modes of
a triaxial nanoellipsoid with $n=1$ at $a_{2}/a_{1}=0.6$ and $%
a_{3}/a_{1}=0.4 $. Index $m=1,2,3$ denotes upper index of the potential
function $\phi _{1}^{\left( m\right) }$. Warm colors correspond to the
positive charge; cold - negative.

Fig.4 (Color online) Relative frequency of plasmon oscillations of a
triaxial nanoellipsoid $\omega _{1}^{\left( m\right) }/\omega _{pl}$ as
function of semi-axes ratio $a_{3}/a_{1}$ at $a_{2}/a_{1}=0.6$.

Fig.5 (Color online) Distribution of the squared electric field logarithm
(a.u.) near a triaxial nanoellipsoid with $a_{2}/a_{1}=0.6$ and $%
a_{3}/a_{1}=0.4$ with semi-axes $a_{2}/a_{1}=0.6$ and $a_{3}/a_{1}=0.4$ in
the plane $x_{3}=0$ for plasmonic modes with $n=2$. Index $m=1,2,\ldots ,5$
denotes upper index of the potential function $\phi _{2}^{(m)}$. Transition
from red color to blue corresponds to transition from the highest squared
values of electric field to the least squared values. Black line corresponds
to the surface of the nanoparticle.

Fig.6 (Color online) Surface charge distribution (a.u.) of plasmon modes of
a triaxial nanoellipsoid with $n=2$ at $a_{2}/a_{1}=0.6$ and $%
a_{3}/a_{1}=0.4 $. Index $m=1,2,\ldots ,5$ denotes upper index of the
potential function $\phi _{2}^{(m)}$. Warm colors correspond to the positive
charge; cold - negative.

Fig.7 (Color online) Relative frequency of plasmonic oscillations of the
triaxial nanoellipsoid $\omega _{2}^{\left( m\right) }/\omega _{pl}$ as the
function of semi-axes ratio $a_{3}/a_{1}$ at $a_{2}/a_{1}=0.6$.

Fig.8 (Color online) Distribution of the squared electric field logarithm
(a.u.) near a triaxial nanoellipsoid with semi-axes $a_{2}/a_{1}=0.6$ and $%
a_{3}/a_{1}=0.4$ in the plane $x_{3}=0$ for plasmon modes with $n=3$. Index $%
m=1,2,\ldots ,7$ denotes upper index of the potential function $\phi
_{3}^{(m)}$. Transition from red color to blue corresponds to transition
from the highest squared values of electric field to the least squared
values. Black line corresponds to the surface of the nanoparticle.

Fig.9 (Color online) Surface charge distribution (a.u.) of plasmon modes of
a triaxial nanoellipsoid with $n=3$ at $a_{2}/a_{1}=0.6$ and $%
a_{3}/a_{1}=0.4 $. Index $m=1,2,\ldots ,7$ denotes upper index of the
potential function $\phi _{3}^{(m)}$. Warm colors correspond to the positive
charge; cold - negative.

Fig.10 (Color online) Relative frequency of the triaxial nanoellipsoid
plasmon oscillations $\omega _{3}^{(m)} / \omega _{pl}$ as function of the
semi-axes ratio $a_{3}/a_{1}$ at $a_{2}/a_{1}=0.6$.

Fig.11 (Color online) Relative frequency of the triaxial nanoellipsoid
plasmon oscillations $\omega _{4}^{(m)} / \omega _{pl} $ as function of the
semi-axes ratio $a_{3}/a_{1}$ at $a_{2}/a_{1}=0.6$.

Fig.12 (Color online) Permittivity of the triaxial nanoellipsoid
plasmon oscillations as function of the semi-axes ratio $a_{3}/a_{1}$ at $%
a_{2}/a_{1}=0.6$. Solid lines represent analytical solution for
$n=1$ (blue lines), $n=2$ (green lines) and $n=3$ (red lines).
Circles show results of numerical calculations within boundary
element method.

\newpage
\begin{figure}[tbp]
\centering \includegraphics[width=6cm,angle=0]{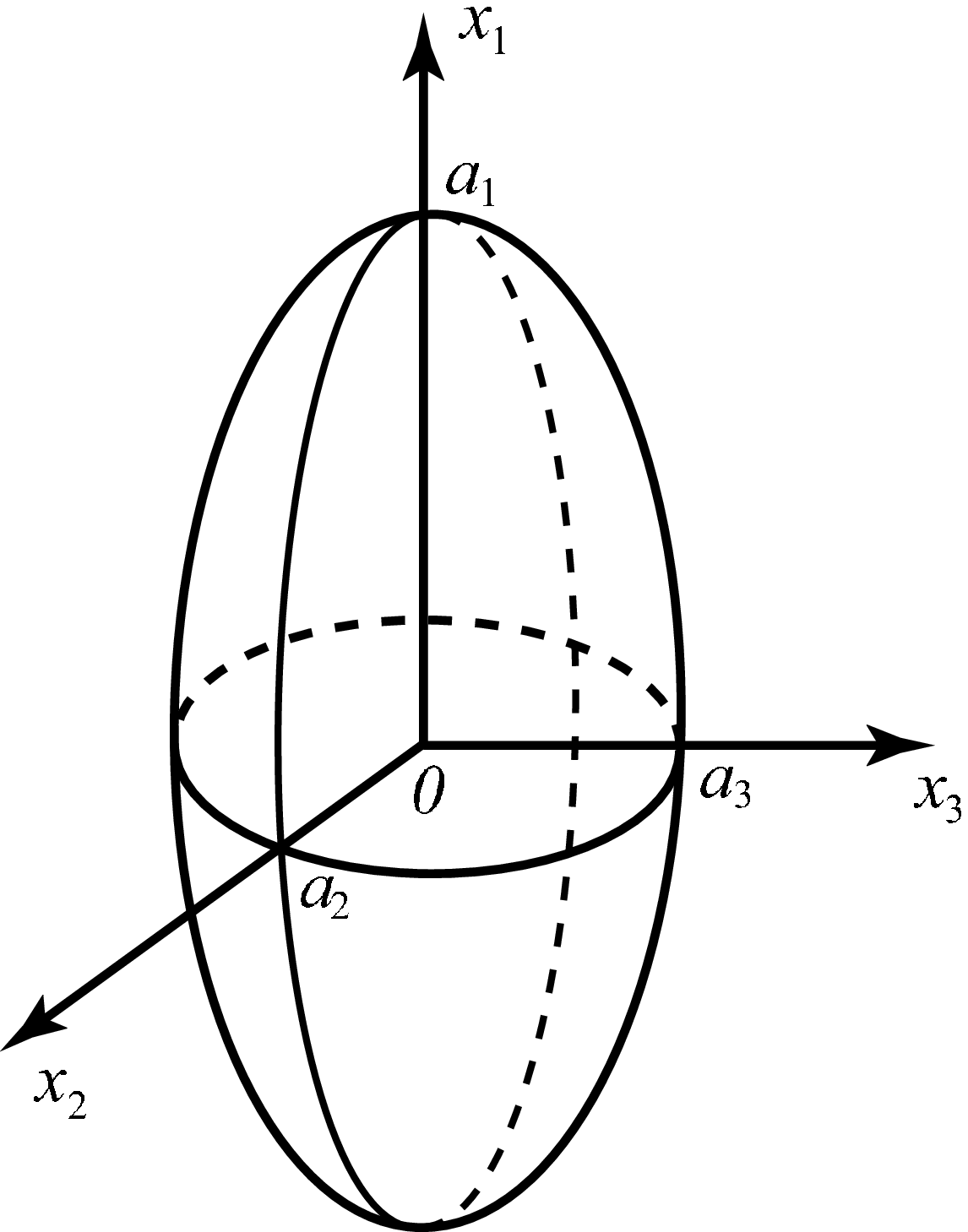} \caption{}
\end{figure}
\pagebreak

\newpage
\begin{figure}[tbp]
\centering \includegraphics[width=6cm,angle=0]{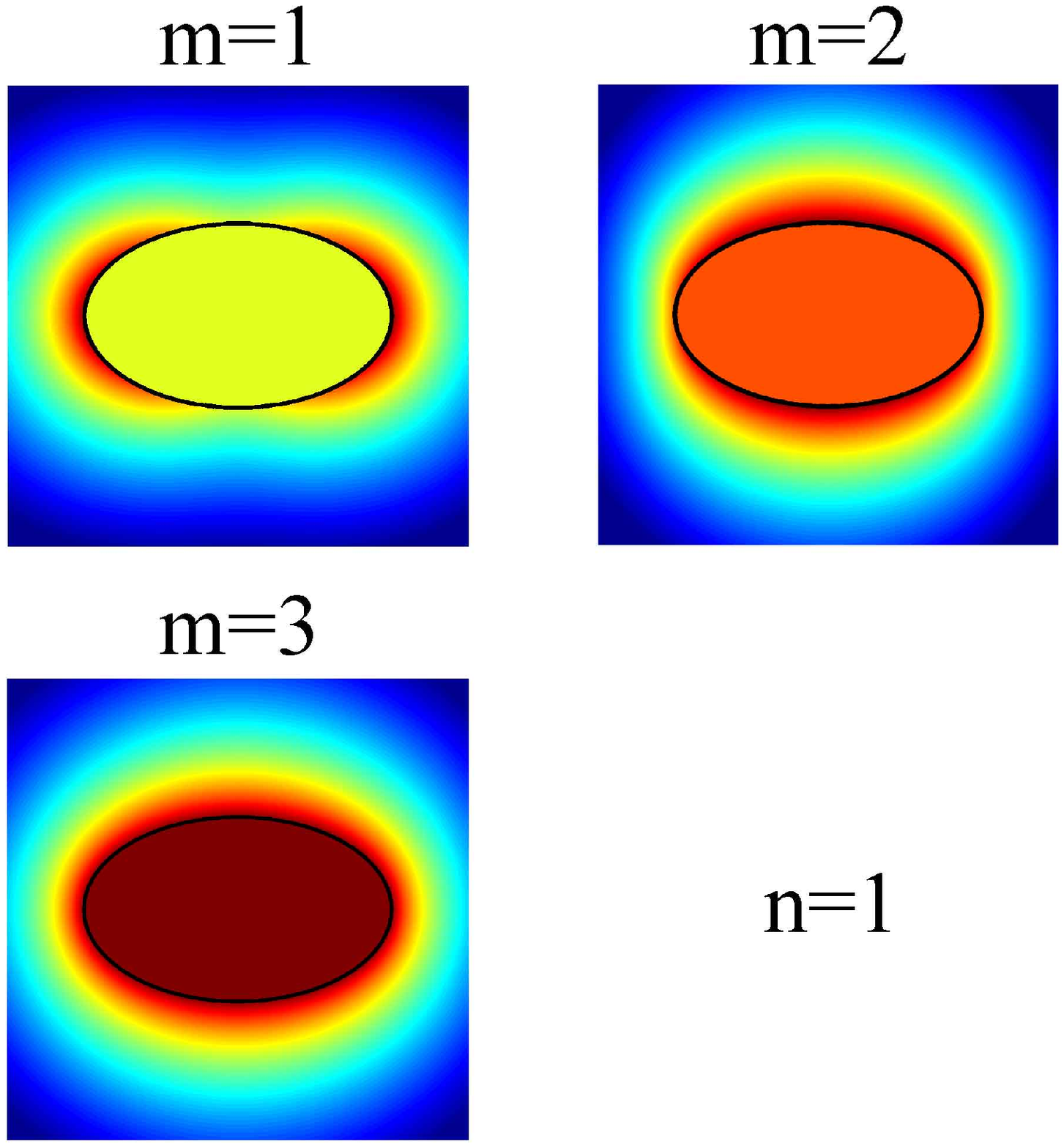} \caption{}
\end{figure}
\pagebreak

\newpage
\begin{figure}[tbp]
\centering \includegraphics[width=6cm,angle=0]{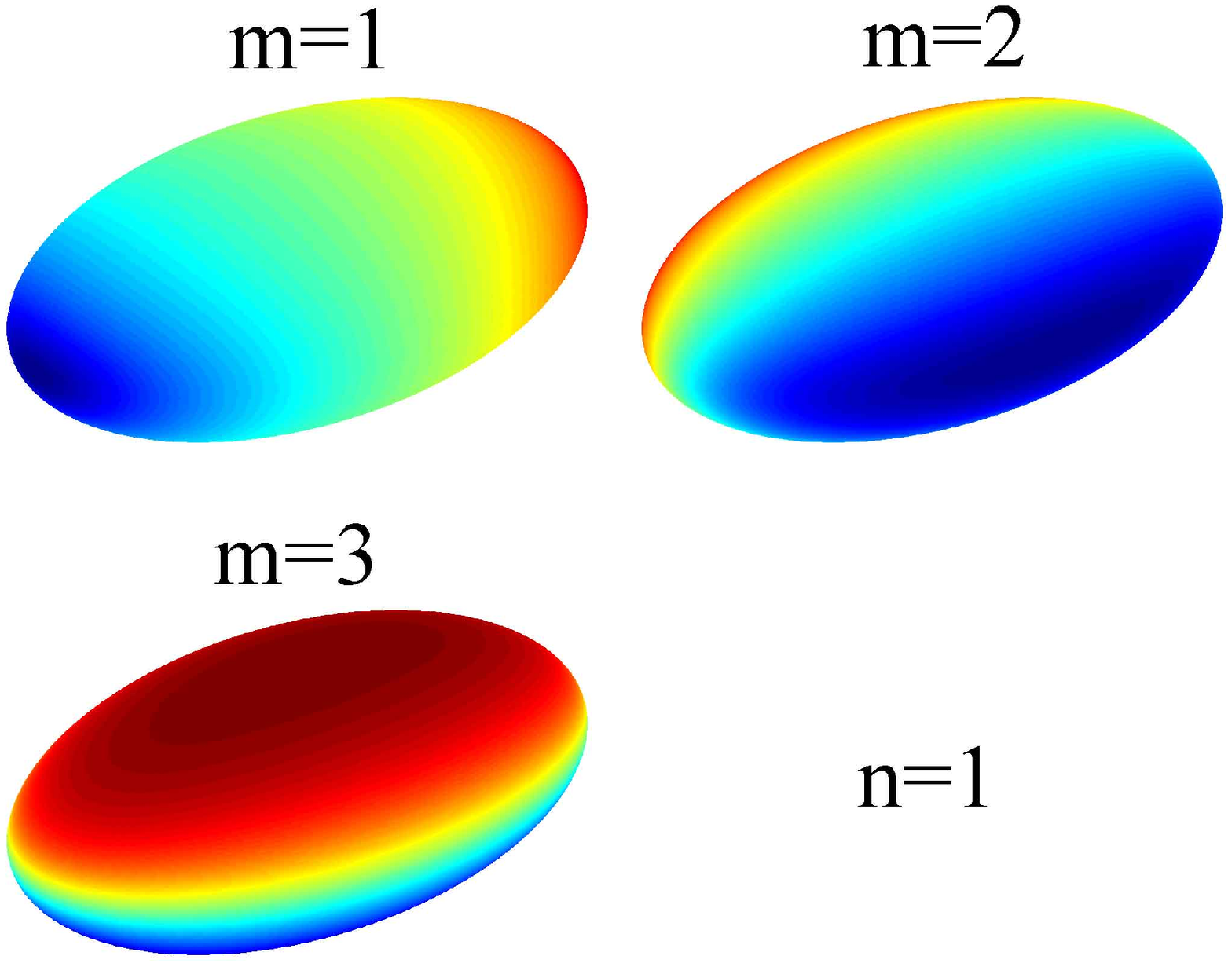} \caption{}
\end{figure}
\pagebreak

\newpage
\begin{figure}[tbp]
\centering \includegraphics[width=9cm,angle=0]{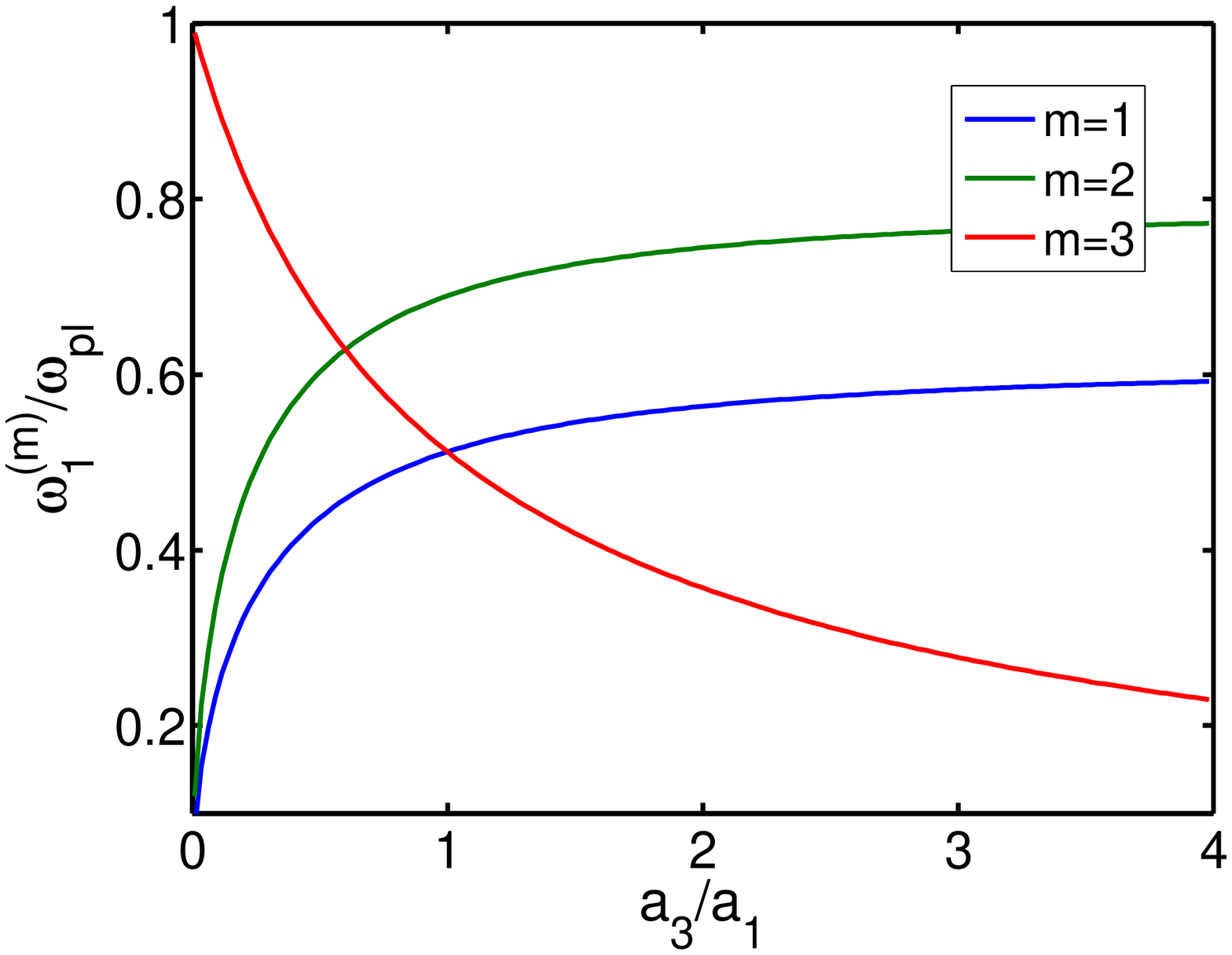} \caption{}
\end{figure}
\pagebreak

\newpage
\begin{figure}[tbp]
\centering \includegraphics[width=6cm,angle=0]{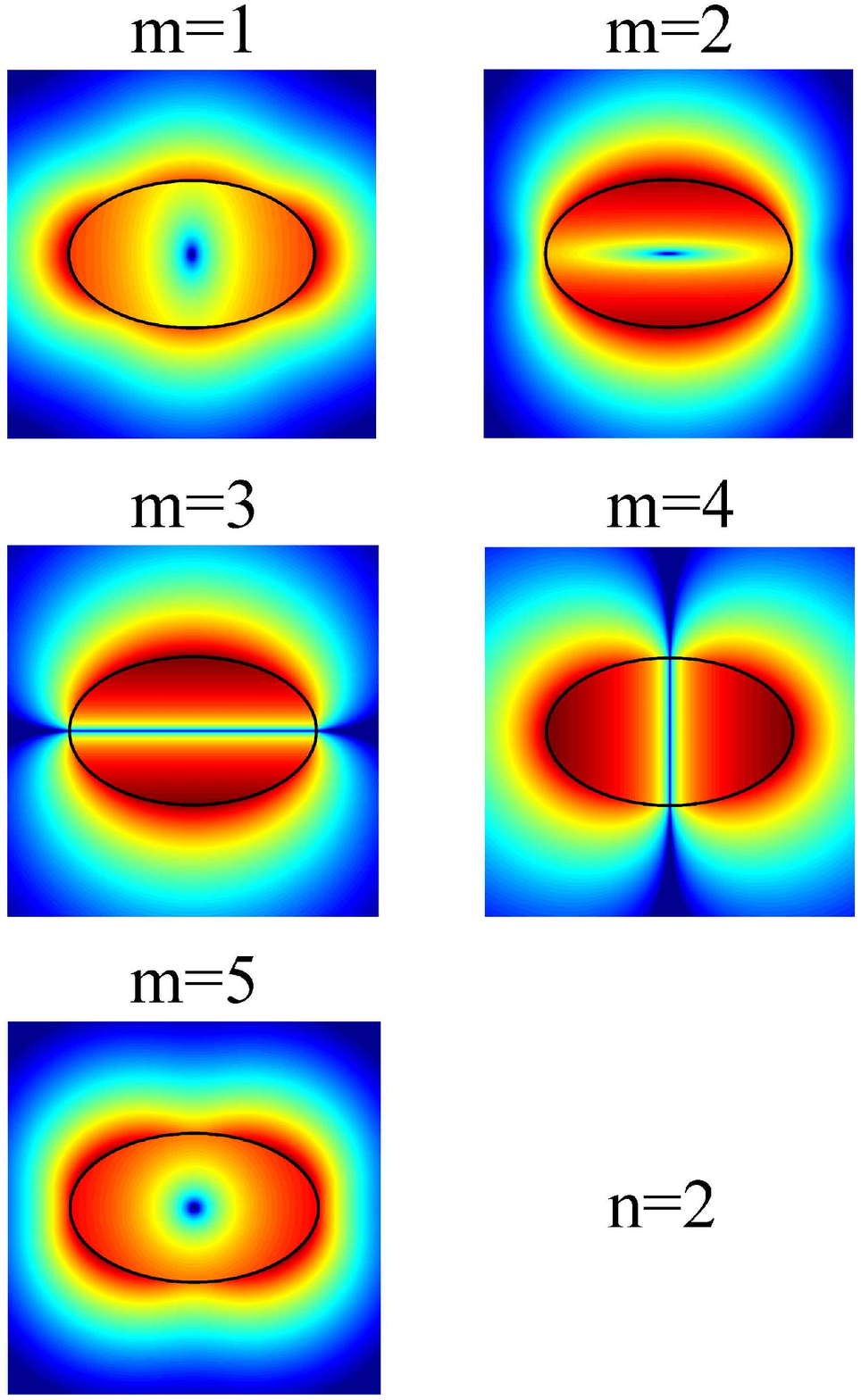} \caption{}
\end{figure}
\pagebreak

\newpage
\begin{figure}[tbp]
\centering \includegraphics[width=6cm,angle=0]{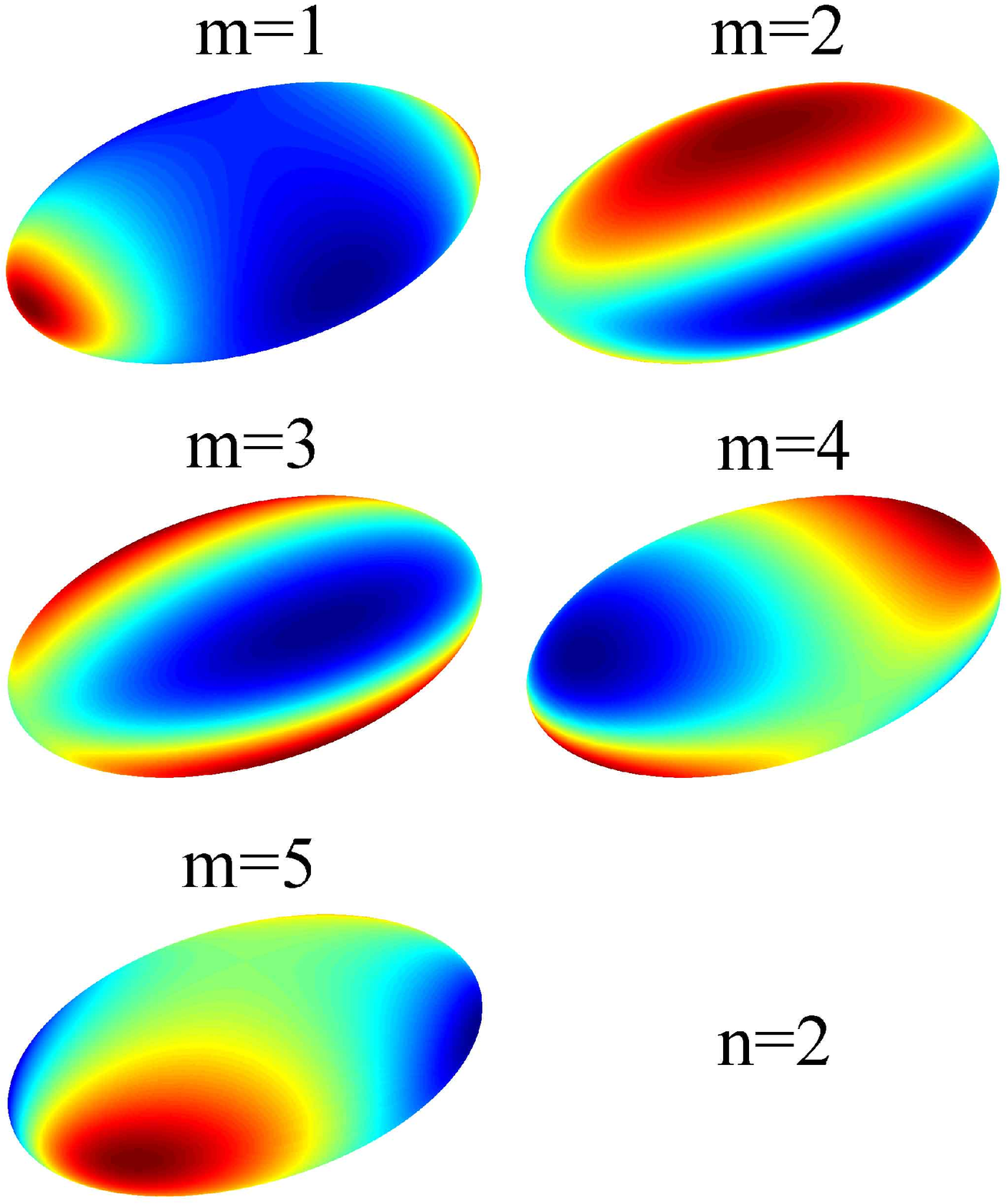} \caption{}
\end{figure}
\pagebreak

\newpage
\begin{figure}[tbp]
\centering \includegraphics[width=9cm,angle=0]{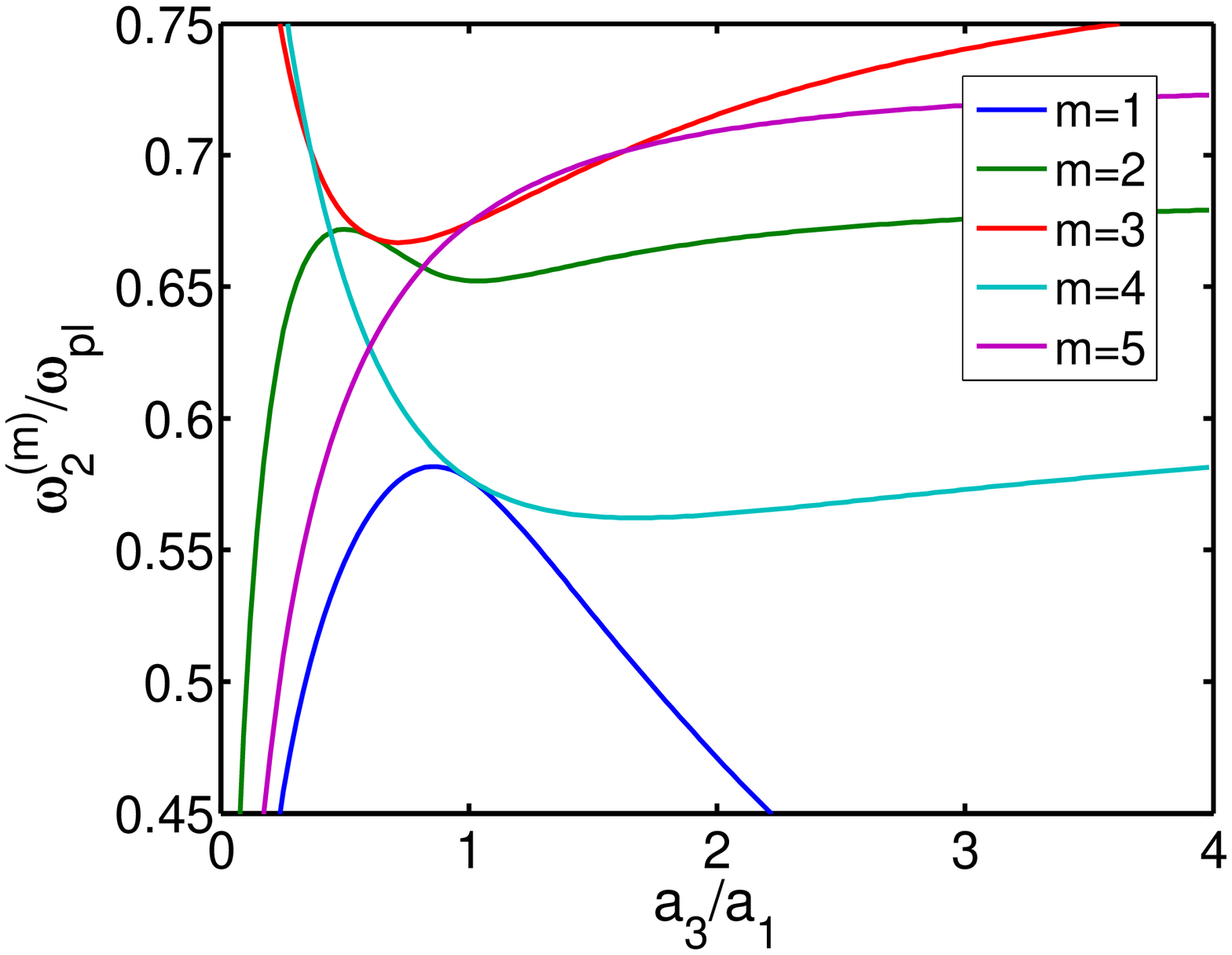} \caption{}
\end{figure}
\pagebreak

\newpage
\begin{figure}[tbp]
\centering \includegraphics[width=6cm,angle=0]{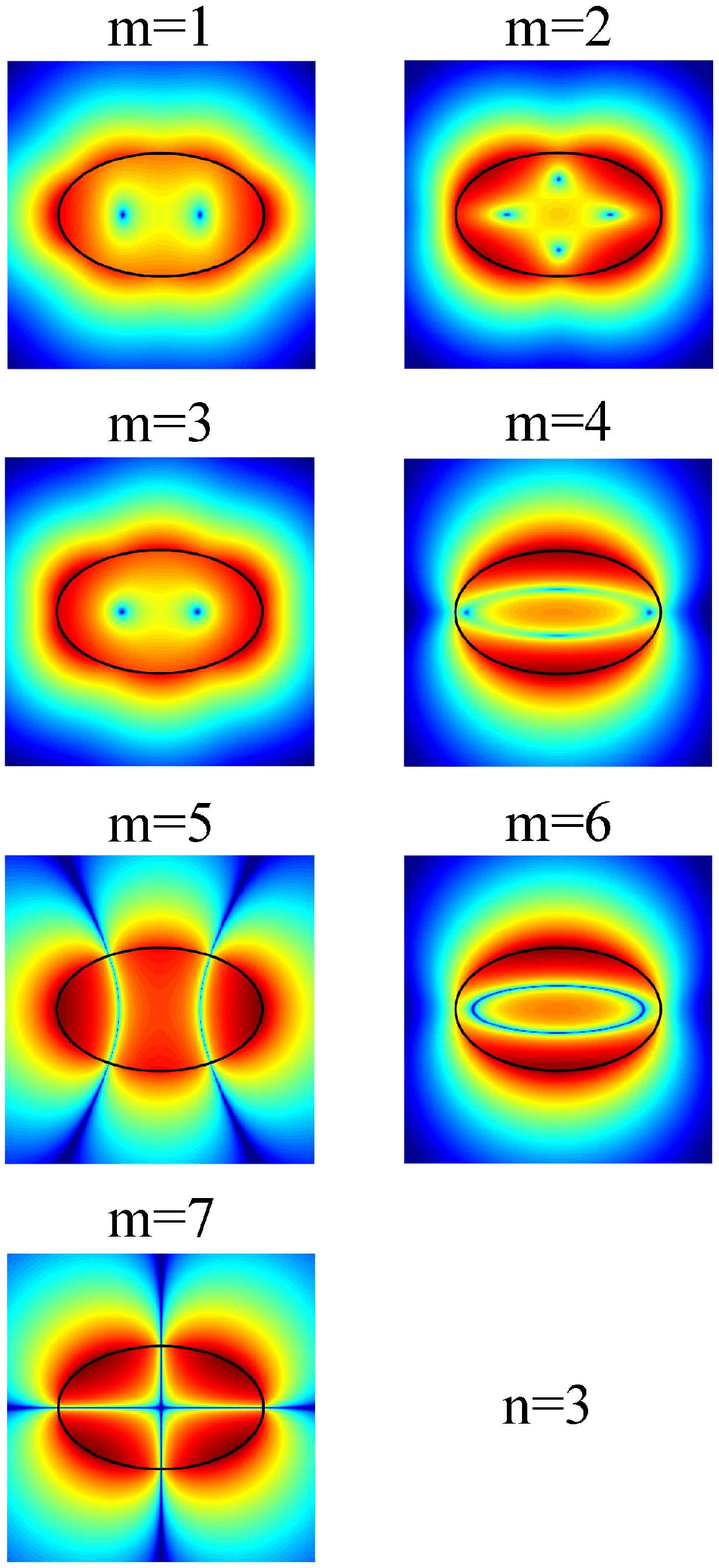} \caption{}
\end{figure}
\pagebreak

\newpage
\begin{figure}[tbp]
\centering \includegraphics[width=6cm,angle=0]{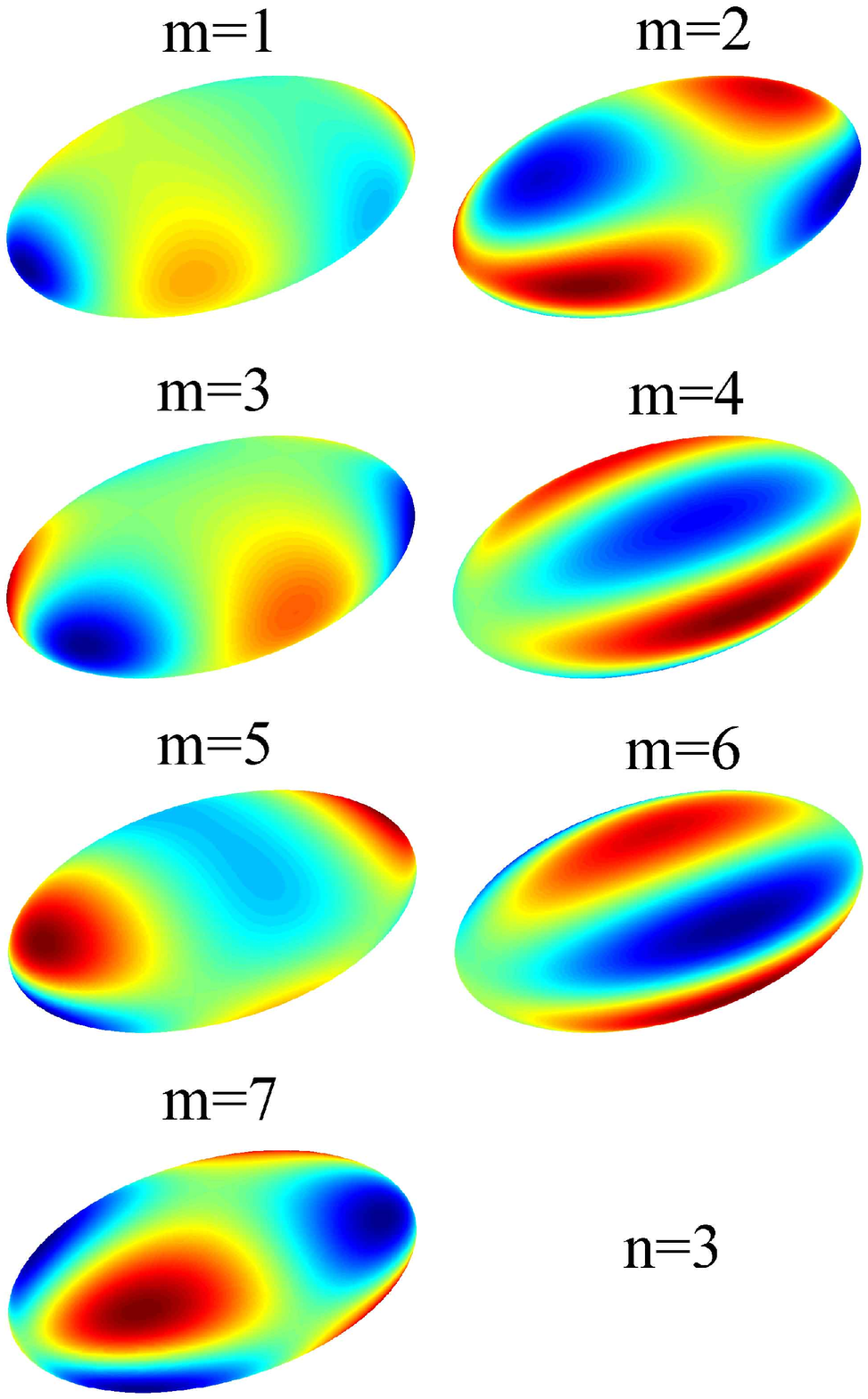} \caption{}
\end{figure}
\pagebreak

\newpage
\begin{figure}[tbp]
\centering \includegraphics[width=9cm,angle=0]{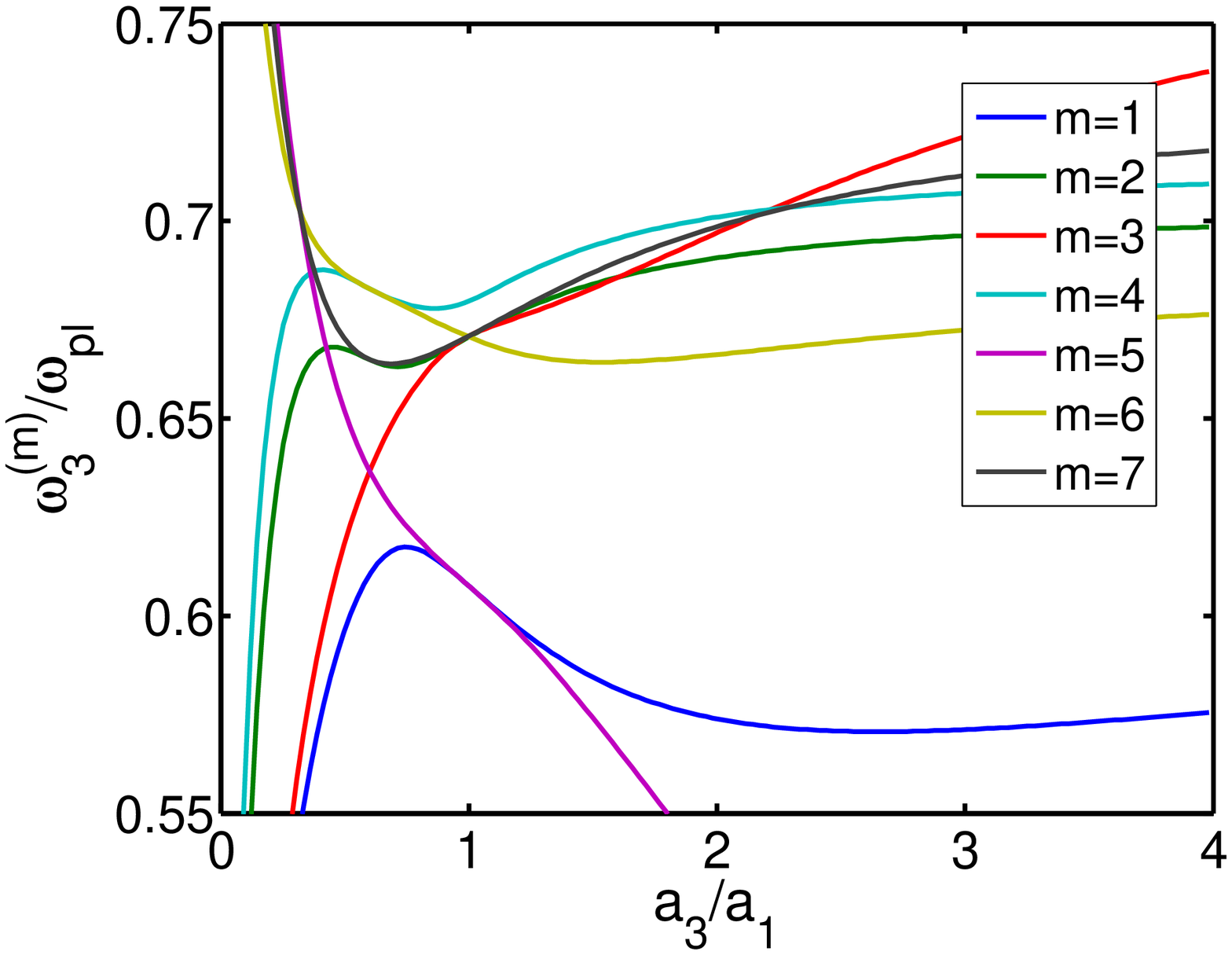} \caption{}
\end{figure}
\pagebreak

\newpage
\begin{figure}[tbp]
\centering \includegraphics[width=9cm,angle=0]{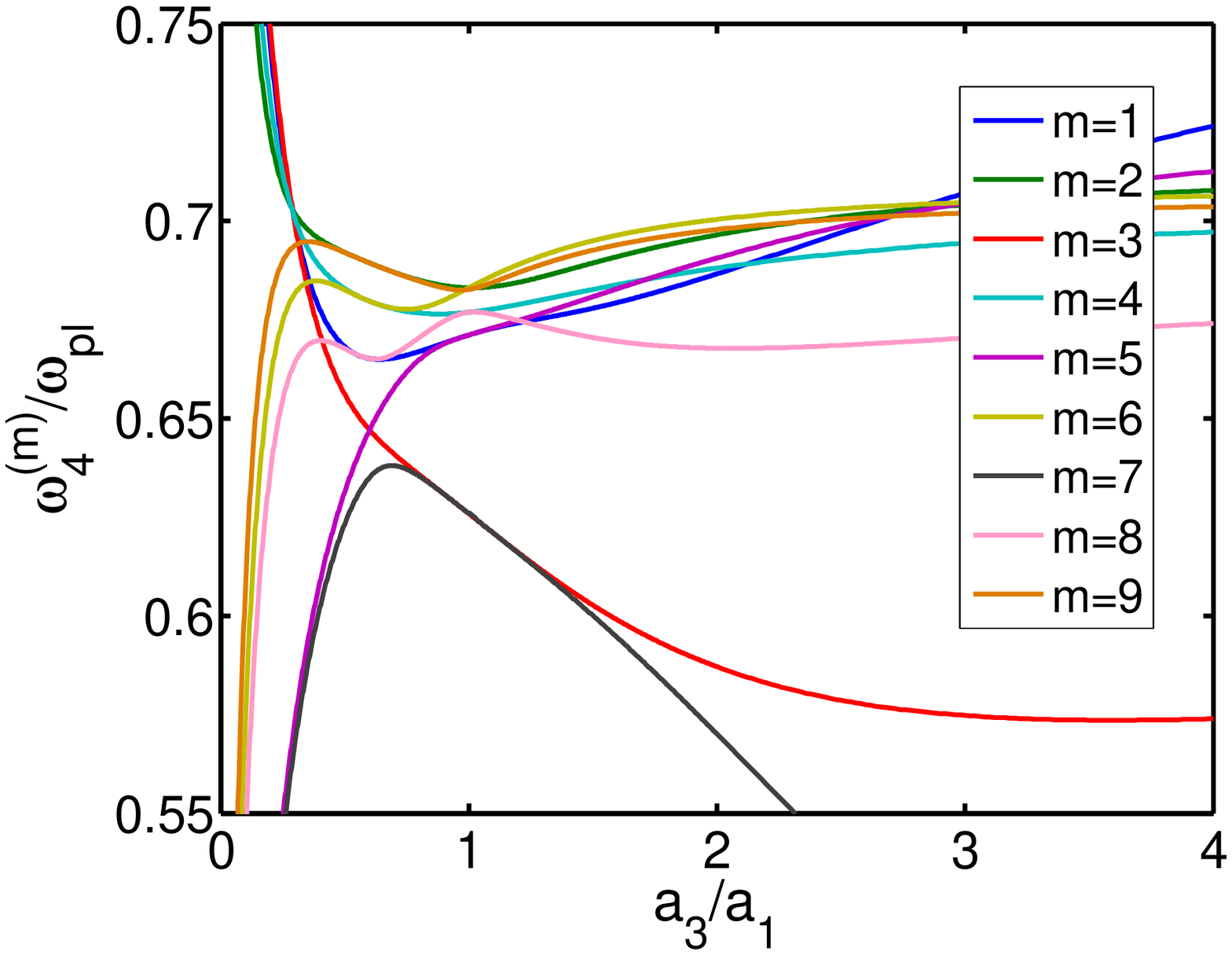} \caption{}
\end{figure}
\pagebreak

\newpage
\begin{figure}[tbp]
\centering \includegraphics[width=9cm,angle=0]{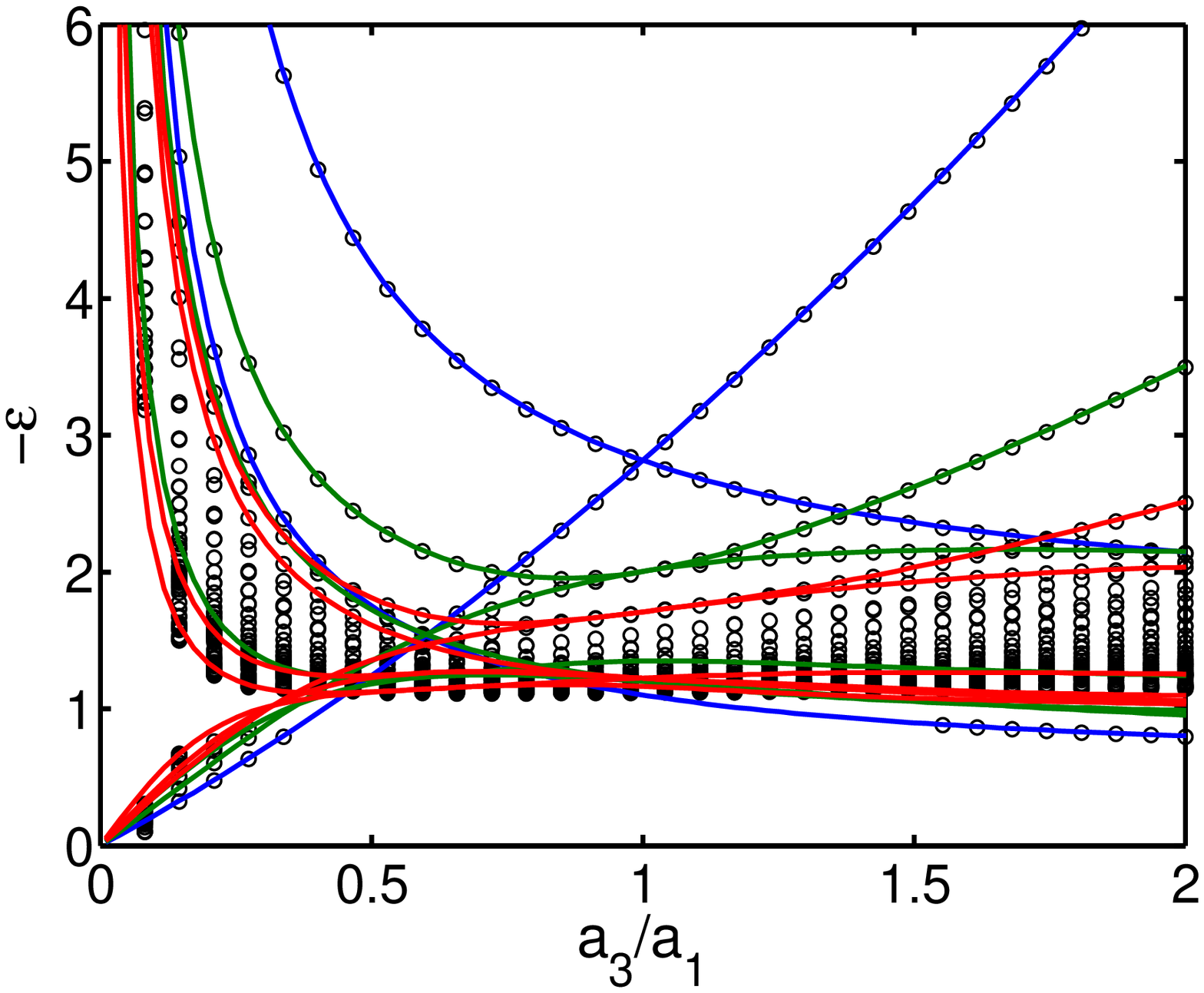} \caption{}
\end{figure}
\pagebreak

\end{document}